\documentclass[sigconf,nonacm]{acmart}

\usepackage{subfigure}
\usepackage{siunitx}
\usepackage{enumitem}

\usepackage{algorithm}
\usepackage[noend]{algpseudocode}
\let\ReturnInline\Return
\renewcommand{\Return}{\State\ReturnInline}
\algrenewcommand\algorithmicrequire{$\rhd$}
\algrenewcommand\algorithmicensure{$\square$}

\AtBeginDocument{%
  \providecommand\BibTeX{{%
    \normalfont B\kern-0.5em{\scshape i\kern-0.25em b}\kern-0.8em\TeX}}}

\setcopyright{acmcopyright}
\copyrightyear{2018}
\acmYear{2018}
\acmDOI{XXXXXXX.XXXXXXX}

\acmConference[Conference acronym 'XX]{Make sure to enter the correct
  conference title from your rights confirmation emai}{June 03--05,
  2018}{Woodstock, NY}

\newcommand{\su}[1]{{{\color{red} #1}}}

\newcommand{\ignore}[1]{}

\begin{document}

\title[Heuristic-based Dynamic Leiden Algorithm for Efficient Tracking of Communities on Evolving Graphs]{Heuristic-based Dynamic Leiden Algorithm for Efficient \\Tracking of Communities on Evolving Graphs}


\author{Subhajit Sahu}
\email{subhajit.sahu@research.iiit.ac.in}
\affiliation{%
  \institution{IIIT Hyderabad}
  \streetaddress{Professor CR Rao Rd, Gachibowli}
  \city{Hyderabad}
  \state{Telangana}
  \country{India}
  \postcode{500032}
}


\settopmatter{printfolios=true}

\begin{abstract}
Community detection, or clustering, identifies groups of nodes in a graph that are more densely connected to each other than to the rest of the network. Given the size and dynamic nature of real-world graphs, efficient community detection is crucial for tracking evolving communities, enhancing our understanding and management of complex systems\ignore{ through real-time analysis}. The Leiden algorithm, which improves upon the Louvain algorithm, efficiently detects communities in large networks, producing high-quality structures. However, existing multicore dynamic community detection algorithms based on Leiden are inefficient and lack support for tracking evolving communities. This technical report introduces the first implementations of parallel Naive-dynamic (ND), Delta-screening (DS), and Dynamic Frontier (DF) Leiden algorithms that efficiently track communities over time. Experiments on a 64-core AMD EPYC-7742 processor demonstrate that ND, DS, and DF Leiden achieve average speedups of $3.9\times$, $4.4\times$, and $6.1\times$, respectively, on large graphs with random batch updates compared to the Static Leiden algorithm, and these approaches scale at $1.4 - 1.5\times$ for every thread doubling.
\end{abstract}



\maketitle

\section{Introduction}
\label{sec:introduction}
Research in graph-structured data has experienced rapid growth due to the capacity of graphs to represent complex, real-world information and capture intricate relationships between entities. A central focus of this field is community detection, which involves dividing a graph into densely interconnected groups, thereby revealing the natural structure within the data. This technique has been applied in a wide range of areas, including: uncovering hidden communities in social networks \cite{blekanov2021detection, la2022information}, examining linguistic variations in memes \cite{zhou2023social}, characterizing polarized information ecosystems \cite{uyheng2021mainstream}, detecting disinformation networks on Telegram \cite{la2021uncovering}, analyzing Twitter communities during the 2022 Ukraine war \cite{sliwa2024case}, analyzing restored Twitter accounts \cite{kapoor2021ll}, reconstructing multi-step cyberattacks \cite{zang2023attack}, developing cyber resilient systems through the study of defense techniques \cite{chernikova2022cyber}, detecting attacks in blockchain systems \cite{erfan2023community}, partitioning large graphs for machine learning \cite{bai2024leiden}, automating microservice decomposition \cite{cao2022implementation}, analyzing regional retail patterns \cite{verhetsel2022regional}, identifying transportation trends \cite{chen2023deciphering}, studying the eco-epidemiology of zoonoses \cite{desvars2024one}, mapping healthcare service areas \cite{wang2021network}, and exploring biological processes \cite{heumos2023best, liu2024sclega, hartman2024peptide, muller2024spatialleiden}.

A key challenge in community detection lies in the lack of prior knowledge about the number of communities and their size distribution. To address this issue, researchers have developed various heuristics for identifying communities \cite{com-blondel08, com-gregory10, com-raghavan07, com-newman16, com-ghoshal19}. The quality of the detected communities is typically evaluated using metrics such as the modularity score introduced by Newman et al. \cite{com-newman04}.

The Louvain method, proposed by Blondel et al. \cite{com-blondel08}, is a widely used community detection algorithm \cite{com-lancichinetti09}. This greedy algorithm uses a two-step approach, consisting of an iterative local-moving phase and an aggregation phase, to iteratively optimize the modularity metric over multiple passes \cite{com-blondel08}. However, Traag et al. \cite{com-traag19} observed that it identify communities that are not only poorly connect, but also internally disconnected. To address this, they propose the \textbf{Leiden algorithm} which adds a refinement phase between the local-moving and aggregation phases. In this refinement phase, vertices can explore and potentially establish sub-communities within those identified during the local-moving phase, enabling the algorithm to better recognize well-connected communities \cite{com-traag19}.

Still, many real-world graphs are vast and evolve rapidly over time. On such graphs, identifying communities in each snapshot of the graph, using \textit{static algorithms} such as the Leiden algorithm, can get quite expensive, both in terms of server running costs and business productivity. In addition, due to the heuristic nature of the algorithms, the obtained communities are inherently unstable, and do not favor tracking the evolution of communities over time.

This is where \textbf{dynamic algorithms} come in. These algorithms can update communities on evolving graphs, without requiring a complete recomputation from scratch. In addition, these algorithms can enable the tracking of community evolution, helping identify key events such as growth, shrinkage, merging, splitting, birth, and death. However, despite the advancements in this area, research has predominantly concentrated on detecting communities in dynamic networks utilizing the Louvain algorithm. We recently proposed three dynamic algorithms \cite{sahu2024starting} based on combining Naive-dynamic (ND) \cite{com-aynaud10}, Delta-screening (DS) \cite{com-zarayeneh21}, and Dynamic Frontier (DF) \cite{sahu2024shared} approaches with one of the most efficient multicore implementations of the Leiden algorithm \cite{sahu2024fast}. Nevertheless, these algorithms leave more to be desired in terms of performance. In addition, we observe that these algorithms are not stable, indicating that the algorithms would likely fail to track communities. In this technical report, we present three techniques/heuristics to improve the performance and stability of the dynamic algorithms.\footnote{\url{https://github.com/puzzlef/leiden-communities-openmp-heuristic-dynamic}}

\section{Related work}
\label{sec:related}
One simple approach to dynamic community detection, referred to as \textbf{Naive-dynamic (ND)}, is to utilize the community memberships of each vertex from the previous graph snapshot, rather than starting with each vertex in its own singleton community \cite{com-aynaud10, com-chong13, com-shang14, com-zhuang19}. More advanced strategies further minimize computational demands by focusing on a smaller subset of the graph that is affected by changes. These strategies include updating only the vertices that have changed \cite{com-aktunc15, com-yin16}, processing vertices near updated edges within a certain threshold distance \cite{com-held16}, breaking down affected communities into lower-level networks \cite{com-cordeiro16}, or using a dynamic modularity metric to recalculate community memberships from scratch \cite{com-meng16}.

\textbf{Delta-screening (DS)} \cite{com-zarayeneh21} and \textbf{Dynamic Frontier (DF)} \cite{sahu2024shared} are two recently proposed approaches for updating communities in a dynamic graph, using the Louvain algorithm. The DS approach analyzes edge deletions and insertions, utilizing the modularity objective, to identify a subset of vertices and communities likely affected by these changes. Only these subsets are then processed for community state updates. On the other hand, the parallel DF approach first identifies an initial set of affected vertices in response to a batch update, and incrementally expands this set of vertices to region surrounding the active/migrating vertices. A parallel algorithm for the DS approach has also been proposed \cite{sahu2024shared}.

In our recent work \cite{sahu2024starting}, which extends ND, DS, and DF approaches to the Leiden algorithm, we made an interesting observation --- for small batch updates, early convergence of the dynamic algorithm can result in suboptimal communities with low modularity. This is because, refinement of communities results is low level (small) community structures. These communities must be hierarchically merged in order to achieve high level community structures with good modularity scores. However, aggregating communities hierarchically is significantly expensive, and defeats the purpose we set out to achieve, i.e., fast identification of communities on (relatively) small batch updates to the graph.

To address the above issue, we proposed a method for selectively refining communities \cite{sahu2024starting}. Specifically, during the local-moving phase of the Leiden algorithm, if a vertex moves from one community to another, both the source and target communities are flagged for refinement. Additionally, we highlighted that edge deletions or insertions within a community can cause it to split, and such communities should also be marked for refinement. In the refinement phase, only these marked communities undergo further processing. However, we also pointed out that refining communities directly may lead to inconsistencies in community IDs, which can cause the undesirable problem of internally disconnected communities. To prevent this, we applied a subset renumbering technique, where each community is assigned the ID of one of its vertices, and all associated data structures are updated accordingly. Despite this, we noted that this method does not improve performance due to imbalanced workload distribution during the aggregation phase. The imbalance arises because some threads are tasked with aggregating very large communities into super-vertices (since they were not refined), while other threads handle smaller, refined communities. To mitigate this imbalance, we employed a small dynamically scheduled chunk size during the aggregation phase, reducing the workload disparity at the expense of some scheduling overhead. Furthermore, to ensure the refinement phase can effectively split isolated communities, we modified the algorithm to aggregate communities into super-vertices based on the refinement phase, rather than the local-moving phase, as originally proposed \cite{com-traag19}.

However, as previously mentioned, these proposed algorithms \cite{sahu2024starting} exhibit limitations in terms of performance, and lack stability, suggesting potential difficulties in tracking communities effectively. We refer the reader to Section \ref{sec:improving-stability} for details.

\section{Preliminaries}
\label{sec:preliminaries}
Let $G(V, E, w)$ represent an undirected graph, where $V$ is the set of vertices, $E$ is the set of edges, and $w_{ij} = w_{ji}$ denotes the positive weight assigned to each edge. In the case of an unweighted graph, we assume each edge has a unit weight, i.e., $w_{ij} = 1$. The set of neighbors for any vertex $i$ is given by $J_i = \{j\ |\ (i, j) \in E\}$, and the weighted degree of vertex $i$ is defined as $K_i = \sum_{j \in J_i} w_{ij}$. The total number of vertices is denoted by $N = |V|$, the total number of edges by $M = |E|$, and the total sum of edge weights in the graph is represented as $m = \sum_{i,j \in V} w_{ij}/2$.

\subsection{Community detection}
\label{sec:about-communities}

Communities that are identified solely based on the network’s structure, without using external information, are referred to as intrinsic, and they are disjoint if each vertex is assigned to only one community \cite{com-gregory10}. Disjoint community detection seeks to determine a community membership function, $C: V \rightarrow \Gamma$, which assigns each vertex $i \in V$ to a community ID $c \in \Gamma$, where $\Gamma$ represents the set of community IDs. The set of vertices within a community $c$ is denoted as $V_c$, and the community containing a vertex $i$ is denoted as $C_i$. Additionally, the neighbors of vertex $i$ that belong to community $c$ are defined as $J_{i \rightarrow c} = \{ j \ |\ j \in J_i \ \text{and} \ C_j = c \}$. The total weight of the edges between vertex $i$ and its neighbors in community $c$ is $K_{i \rightarrow c} = \{ w_{ij} \ |\ j \in J_{i \rightarrow c} \}$, the sum of edge weights within community $c$ is $\sigma_c = \sum_{(i, j) \in E \ \text{and} \ C_i = C_j = c} w_{ij}$, and the total edge weight associated with $c$ is $\Sigma_c = \sum_{(i, j) \in E \ \text{and} \ C_i = c} w_{ij}$ \cite{com-zarayeneh21}.

\subsection{Modularity}
\label{sec:about-modularity}

Modularity is a metric used to evaluate the quality of communities identified by community detection algorithms, which are often heuristic-based \cite{com-newman04}. Its value ranges from $[-0.5, 1]$ (with higher values indicating better community structure) and is calculated as the difference between the actual fraction of edges within communities and the expected fraction of edges if they were randomly distributed \cite{com-brandes07}. The modularity score, $Q$, for the detected communities is computed using Equation \ref{eq:modularity}. The change in modularity, or \textit{delta modularity}, when moving a vertex $i$ from community $d$ to community $c$ --- denoted as $\Delta Q_{i: d \rightarrow c}$ --- is given by Equation \ref{eq:delta-modularity}.

\begin{equation}
\label{eq:modularity}
  Q
  = \sum_{c \in \Gamma} \left[\frac{\sigma_c}{2m} - \left(\frac{\Sigma_c}{2m}\right)^2\right]
\end{equation}

\begin{equation}
\label{eq:delta-modularity}
  \Delta Q_{i: d \rightarrow c}
  = \frac{1}{m} (K_{i \rightarrow c} - K_{i \rightarrow d}) - \frac{K_i}{2m^2} (K_i + \Sigma_c - \Sigma_d)
\end{equation}

\subsection{Louvain algorithm}
\label{sec:about-louvain}

The Louvain method is a greedy, modularity optimization based agglomerative algorithm for detecting communities of high quality in a graph. It works in two phases: in the local-moving phase, each vertex $i$ considers moving to a neighboring community $C_j\ |\ j \in J_i$ to maximize modularity increase $\Delta Q_{i:C_i \rightarrow C_j}$. In the aggregation phase, vertices in the same community are merged into super-vertices. These phases constitute a pass, and repeat until modularity gain stops. This produces a hierarchical structure (dendrogram), with the top level representing the final communities.

\subsection{Leiden algorithm}
\label{sec:about-leiden}

The Leiden algorithm, proposed by Traag et al. \cite{com-traag19}, addresses the connectivity issues of the Louvain method by adding a refinement phase after the local-moving phase. In this \textit{refinement phase}, vertices in each community perform constrained merges into neighboring sub-communities, starting from singleton sub-communities established earlier. The merging process is randomized, with a vertex's probability of joining a neighboring sub-community proportional to the delta-modularity of the move. This approach enhances the identification of well-separated and well-connected sub-communities. The algorithm has a time complexity of $O(L|E|)$, where $L$ is the total number of iterations, and a space complexity of $O(|V| + |E|)$, similar to the Louvain method.

\subsection{Dynamic approaches}
\label{sec:dynamic-graphs}

A dynamic graph is represented as a sequence of graphs, denoted by $G^t(V^t, E^t, w^t)$ at time step $t$. Changes between consecutive graphs $G^{t-1}(V^{t-1}, E^{t-1}, w^{t-1})$ and $G^t(V^t, E^t, w^t)$ are captured in a \textit{batch update} $\Delta^t$ at time $t$. This\ignore{update} includes edge deletions $\Delta^{t-} = E^{t-1} \setminus E^t$ and edge insertions $\Delta^{t+} = \{(i, j, w_{ij})\ |\ w_{ij} > 0\} = E^t \setminus E^{t-1}$\ignore{\cite{com-zarayeneh21}}.

\subsubsection{Naive-dynamic (ND) approach \cite{com-aynaud10, com-chong13, com-shang14, com-zhuang19}}
\label{sec:about-naive}

The Naive dynamic approach is a straightforward method for detecting communities in dynamic networks. Vertices are assigned to communities based on the previous graph snapshot, and all vertices are processed, irrespective of\ignore{any} edge deletions/insertions in the batch update\ignore{, which is why it is termed \textit{naive}}.

\subsubsection{Delta-screening (DS) approach \cite{com-zarayeneh21}}
\label{sec:about-delta}

\begin{figure}[hbtp]
  \centering
  \subfigure[Delta-screening (DS)]{
    \label{fig:about-cases--delta}
    \includegraphics[width=0.468\linewidth]{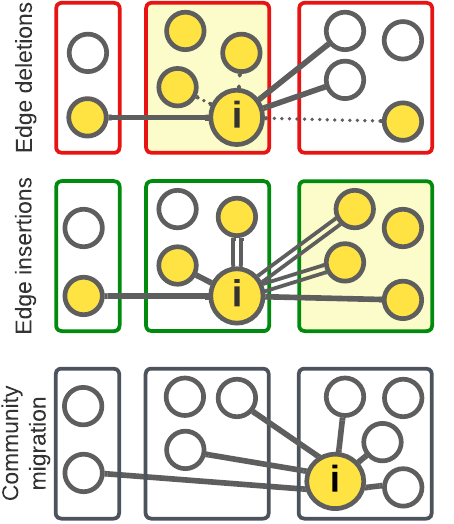}
  }
  \subfigure[Dynamic Frontier (DF)]{
    \label{fig:about-cases--frontier}
    \includegraphics[width=0.425\linewidth]{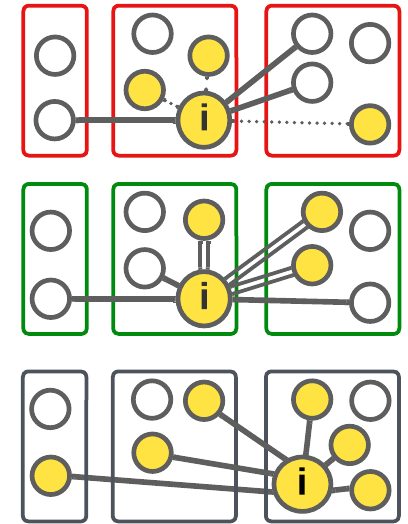}
  } \\[-2ex]
  \caption{Illustration of \textit{Delta-screening (DS)} \cite{com-zarayeneh21} and \textit{Dynamic Frontier (DF)} approaches \cite{sahu2024shared}, in the presence of edge deletions (dotted lines) and insertions (doubled lines). Vertices identified as affected (initially) by each approach are highlighted in yellow, while affected entire communities are shaded in light yellow.}
  \label{fig:about-cases}
\end{figure}

This dynamic community detection method employs modularity-based scoring to pinpoint regions of a graph where vertices may change their community membership. Zarayeneh et al. implement the DS approach by first sorting batch updates of edge deletions $(i, j) \in \Delta^{t-}$ and insertions $(i, j, w) \in \Delta^{t+}$ by source vertex ID. For edge deletions within the same community, they mark the neighbors of vertex $i$ and the community of vertex $j$ as affected. For edge insertions across communities, they identify vertex $j^*$ with the highest modularity change linked to vertex $i$ and mark the neighbors of $i$ and the community of $j^*$ as affected. Edge deletions between different communities and insertions within the same community are considered unlikely to influence community membership and are ignored. Figure \ref{fig:about-cases--delta} illustrates the affected vertices and communities related to a single source vertex $i$ in response to a batch update\ignore{involving both deletions and insertions}.

\subsubsection{Dynamic Frontier (DF) approach \cite{sahu2024shared}}
\label{sec:about-frontier}

This approach begins by initializing each vertex's community membership based on the previous graph snapshot. When edges are deleted between vertices in the same community or inserted between different communities, the source vertex $i$ is marked as affected. Since batch updates are undirected, both endpoints $i$ and $j$ are marked. Edge deletions between different communities or insertions within the same community are ignored\ignore{(as noted in Section \ref{sec:about-delta})}. Additionally, if vertex $i$ changes its community membership, all its neighboring vertices $j \in J_i$ are marked as affected, while $i$ is marked as unaffected. To reduce unnecessary computations, an affected vertex $i$ is also marked as unaffected if its community remains unchanged. The DF approach then employs a graph traversal-like method until the community assignments stabilize. Figure \ref{fig:about-cases--frontier} illustrates the vertices connected to a source vertex $i$ marked as affected by the DF approach after a batch update.

\section{Approach}
\label{sec:approach}
We now discuss\ignore{our} techniques to address the two keys issues with the dynamic algorithms presented earlier by us \cite{sahu2024starting}, i.e., the stability of obtained communities, and performance.

\subsection{Improving stability}
\label{sec:improving-stability}

Let us first consider addressing the issue of stability of obtained communities on dynamic graphs. Here, we are essentially asking the question of how well a dynamic community detection algorithm is able to help us keep track of the identified communities, by continuing to preserve the associated ID for each community across small updates, which modify the vertex set associated with the community, but do not significantly change it --- or if they do, the change is slow, i.e., it appears over a number of batch updates.

\begin{figure}[!hbt]
  \centering
  \subfigure{
    \label{fig:about-address-stability--all}
    \includegraphics[width=0.98\linewidth]{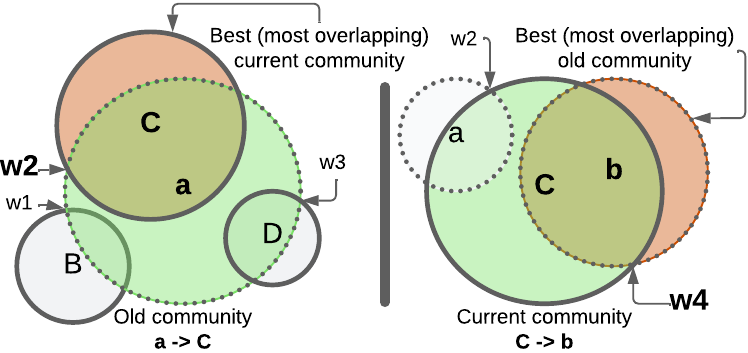}
  } \\[-2ex]
  \caption{Our procedure for tracking of communities. In the figure, old communities are labeled with lowercase characters and have dotted border (e.g., $a$ and $b$), while current communities are labeled with uppercase characters and have a solid border (e.g. $B$, $C$, and $D$). An overlap of vertices between old and current communities is indicated with an intersection, as in a venn diagram, with the total edge weight of the overlaps being indicated with $w1$, $w2$, $w3$, or $w4$.}
  \label{fig:about-address-stability}
\end{figure}

We propose the following procedure for tracking communities. It involves first identifying the most overlapping community $C_i \in \Gamma^t$ (by total edge weight) to each old community $b_i \in \Gamma^{t-1}$ in the previous graph snapshot, and then identifying the most overlapping old community $b_j \in \Gamma^{t-1}$ to each current community $C_j \in \Gamma^t$ (since a current community may be the best overlapping community for multiple old communities) --- to create a mapping $s: \Gamma^t \rightarrow \Gamma^{t-1}$ from each current community ID, $C_j$, to the best matching old community ID, $b_j$. Take a look at Figure \ref{fig:about-address-stability} for an example. As shown in the left subfigure, we first identify that the current community $C$ is the most overlapping community with the old community $a$ (with a total overlap edge weight $w2$). In the right subfigure, we observe that the current community $C$ has best overlaps with two old communities $a$ and $b$, but since $b$ is the most overlapping old community (with a total overlap edge weight $w4$), the final mapping created is from current community $C$ to old community $b$. We use the Boyer-Moore majority vote algorithm \cite{boyer1991mjrty} in order to obtain the most (majority) overlapping community in a reasonable amount of time. We refer the reader to Section \ref{sec:our-tracking} for more details.

We now wish to measure the stability of communities identified with our proposed procedure, using ND, DS, and DF Leiden. For this, we consider the following experiment. First, we obtain the community membership of vertices in the initial snapshot of a graph $G^0$, using Static Leiden. Next, we apply a random batch update, consisting purely of edge deletions, in order to obtain an updated graph $G^1$. Here, we use each dynamic algorithm to obtain the updated community membership of vertices. Finally, we reinsert the same edges back to obtain the original graph $G^2 = G^0$, and reuse each dynamic algorithm to obtain the updated community membership of vertices from the membership obtained in $G^1$. In the ideal case, the community memberships obtained by the dynamic algorithms on $G^2 = G^0$ would have an exact match with the community memberships obtained by Static Leiden on $G^0$. However, this is can be hard to achieve in practice. One of the reasons is that the refined communities may split off, and rejoin back, but will very likely not have the same community ID when they reform\ignore{again}.

\begin{figure}[hbtp]
  \centering
  \subfigure{
    \label{fig:8020-stability--all}
    \includegraphics[width=0.98\linewidth]{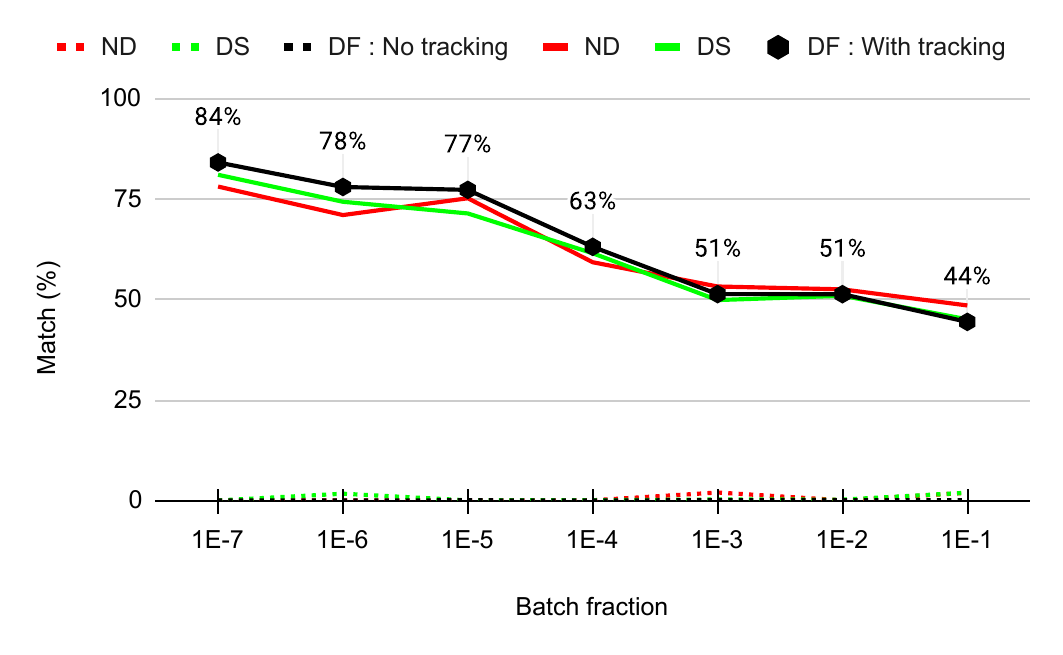}
  } \\[-4ex]
  \caption{Percentage match in the community membership of vertices, after a random batch update of size $10^{-7}|E|$ to $0.1|E|$, consisting purely of edge deletions, followed by a batch update which reverses the edge deletions (by inserting the edges back). The dynamic algorithms compared here include the original \textit{Naive-dynamic (ND)}, \textit{Delta-screening (DS)}, and \textit{Dynamic Frontier (DF) Leiden} with no tracking \cite{sahu2024starting}, and our improved ND, DS, and DF Leiden with tracking. The match $\%$ of our improved DF Leiden is labeled here.}
  \label{fig:8020-stability}
\end{figure}

Figure \ref{fig:8020-stability} shows the percent match in the community membership of vertices with the original ND, DS, and DF Leiden (no tracking) \cite{sahu2024starting}, and our improved ND, DS, and DF Leiden (with tracking). For this experiment, we use graphs from Table \ref{tab:dataset-large}, with the batch size being adjusted from $10^{-7}|E|$ to $0.1|E|$, and the arithmetic mean of the match percentage is plotted. As the figure shows, in the absence of tracking, ND, DS, and DF Leiden have a near zero match with the communities returned by Static Leiden of $G^0$, and are thus not the right candidates for tracking evolving communities on dynamic graphs. In contrast, our improved ND, DS, and DF Leiden\ignore{(with tracking)} achieve a good match, with our improved DF Leiden achieving a match of $84\%$ to $44\%$ on batch updates of size $10^{-7}|E|$ to $0.1|E|$.

\begin{figure}[!hbt]
  \centering
  \subfigure{
    \label{fig:8020-tracking-cost--all}
    \includegraphics[width=0.98\linewidth]{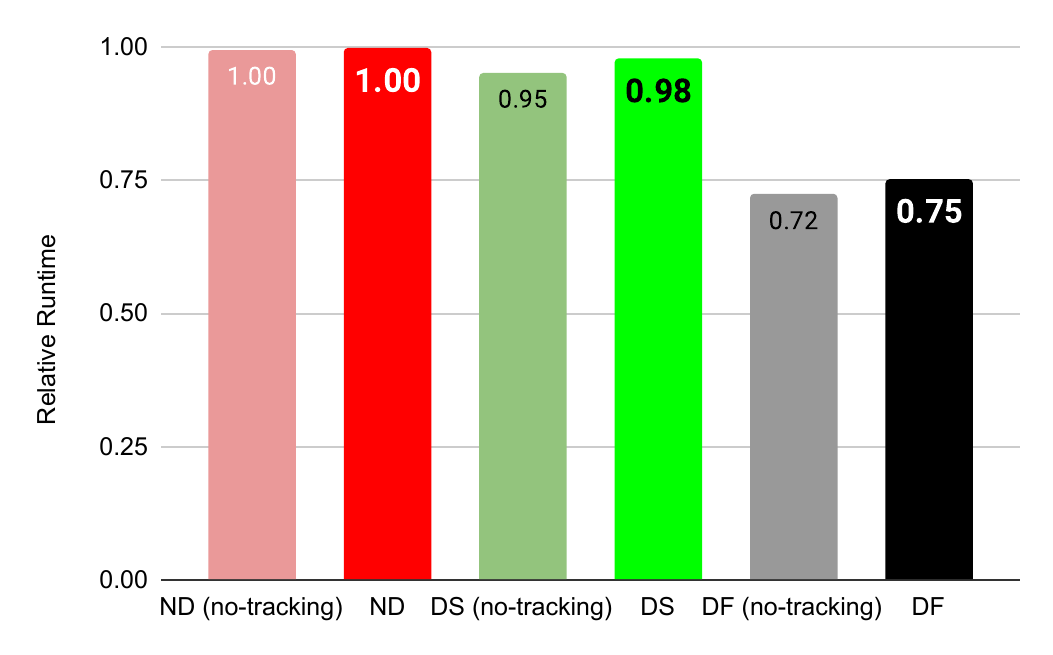}
  } \\[-2ex]
  \caption{Relative Runtime of \textit{Naive-dynamic (ND)}, \textit{Delta-screening (DS)}, and \textit{Dynamic Frontier (DF) Leiden}, evaluated, both with and without community tracking (referred to as "no-tracking"). This experiment was conducted on large graphs with random batch updates of size $10^{-7}|E|$ to $0.1|E|$.}
  \label{fig:8020-tracking-cost}
\end{figure}

Next, we measure the runtime cost of the proposed procedure for tracking of communities --- on large graphs, from Table \ref{tab:dataset-large}, with randomly generated batch updates of size $10^{-7}|E|$ to $0.1|E|$. Figure \ref{fig:8020-tracking-cost} shows the relative runtime of the original ND, DS, and DF Leiden (no-tracking) \cite{sahu2024starting}, compared to our improved ND, DS, and DF Leiden. As the figure shows, the inclusion of the community tracking procedure incurs a minimal additional runtime overhead.

\subsection{Minimizing communities to refine}

We now focus on improving the speed of our dynamic algorithms. In particular, we note that once we refine a community, we must run the algorithm to the end. This means we should aim to minimize the number of communities that require refinement. Note that refinement offers us two key benefits: first, it ensures that we obtain well-connected communities without internal disconnections; second, it allows us to address edge cases, where an isolated community can independently split due to edge deletions or insertions. However, such edge case occurrences are relatively rare.

\begin{figure}[hbtp]
  \centering
  \subfigure{
    \label{fig:about-address-refine--01}
    \includegraphics[width=0.98\linewidth]{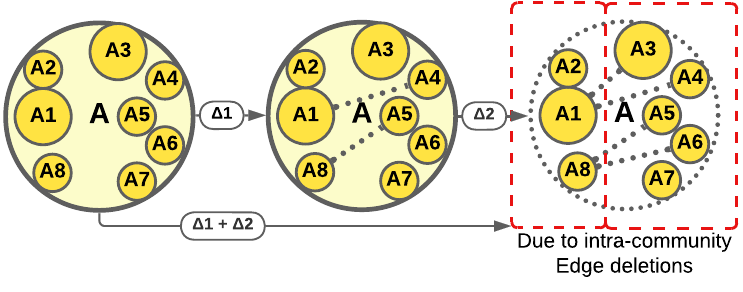}
  } \\[-2ex]
  \subfigure{
    \label{fig:about-address-refine--02}
    \includegraphics[width=0.98\linewidth]{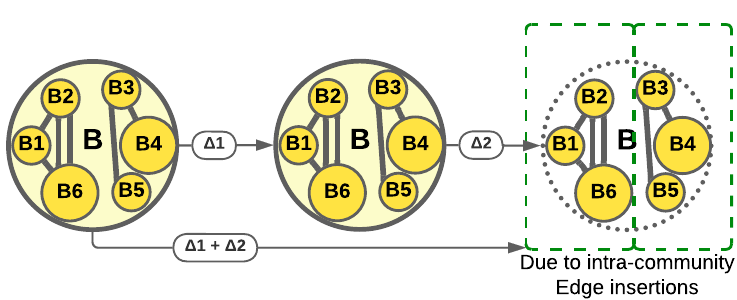}
  } \\[-2ex]
  \caption{Heuristic for minimizing communities to refine. Communities $A$ and $B$ contain subcommunities (labeled $A1$ to $A7$ and $B1$ to $B6$). The top subfigure illustrates that edge deletions (dotted lines) in community $A$ can lead to its split, necessitating refinement. The bottom subfigure shows that edge insertions (solid lines) in $B$ can similarly cause a split due to stronger regional connections, also requiring refinement. The figure demonstrates that our algorithm effectively processes two cumulative batch updates, $\Delta1$ and $\Delta2$, as if they were a single large update $\Delta1 + \Delta2$.}
  \label{fig:about-address-refine}
\end{figure}

To optimize this process, we propose the following heuristic: we will mark a community for refinement only if the cumulative change in its total edge weight changes by a fraction greater than the refinement tolerance $\tau_{re}$, considering both edge deletions and insertions as positive changes to community weight. Since the change in total edge weight of each community is aggregated until a community is marked for refinement, multiple sequential small batch updates have a similar effect as a single large batch update, as shown in Figure \ref{fig:about-address-refine}. When the change in total edge weight of a community exceeds this threshold, we mark it for refinement and reset the current total edge weight change for that community to zero. If no communities are marked for refinement, we can achieve convergence in a single pass, eliminating the need to run the algorithm till the end. The figure also shows how both edge deletions and edge insertions within a community may cause a community to split, and thus a community must be marked for refinement in both cases (to allow for subcommunities to be identified).

\begin{figure}[!hbt]
  \centering
  \subfigure{
    \label{fig:fig-8020-adjust-splitall--all}
    \includegraphics[width=0.98\linewidth]{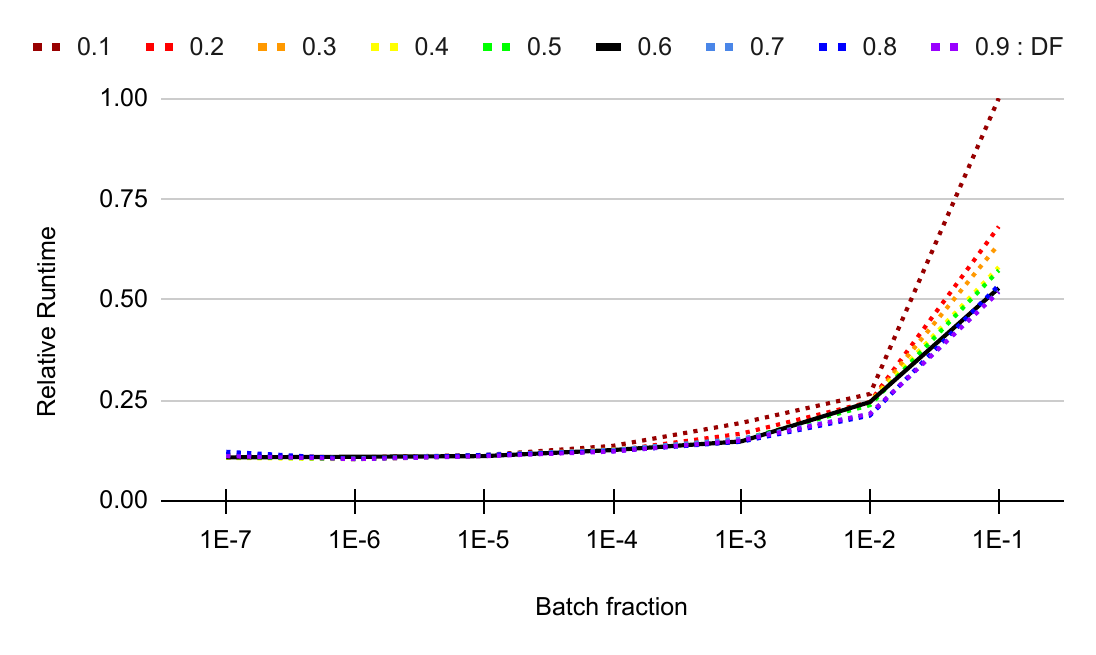}
  } \\[-4ex]
  \caption{Relative Runtime of \textit{Dynamic Frontier (DF) Leiden}, with refinement tolerance $\tau_{re}$ being adjusted from $0.1$ to $0.9$ --- indicating that a community is marked to be refined only if its community weight (considering both edge insertions and deletions\ignore{as a positive change}) cumulatively changes by $10\%$ to $90\%$, respectively.}
  \label{fig:fig-8020-adjust-splitall}
\end{figure}

We now conduct an experiment to measure a suitable value for refinement tolerance $\tau_{re}$. Here, we adjust $\tau_{re}$ from $0.1$ to $0.9$, and test on graphs from Table \ref{tab:dataset-large}, with random batch updates of size $10^{-7}|E|$ to $0.1|E|$. Figure \ref{fig:fig-8020-adjust-splitall} shows the relative runtime of DF Leiden, with $\tau_{re}$ being adjusted. Results suggest a $\tau_{re}$ of $0.6$ to be suitable.

\subsection{Minimizing communities to split}

\begin{figure}[hbtp]
  \centering
  \subfigure{
    \label{fig:about-address-split--01}
    \includegraphics[width=0.88\linewidth]{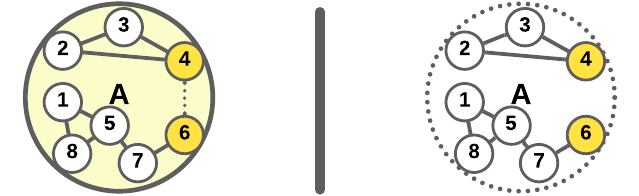}
  } \\[-1ex]
  \subfigure{
    \label{fig:about-address-split--03}
    \includegraphics[width=0.98\linewidth]{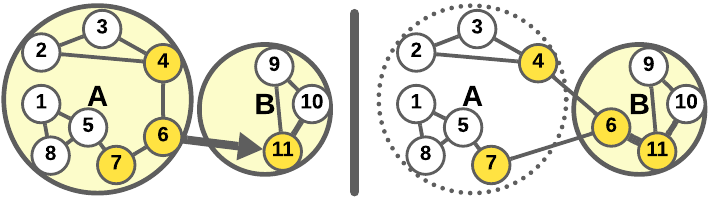}
  } \\[-2ex]
  \caption{Two examples demonstrating situations where a community may become internally disconnected, requiring it to be flagged for splitting. In the top example, community $A$ consists of $8$ vertices. An edge deletion between vertices $4$ and $6$ causes the community to be internally-disconnected. In the bottom example, community $A$ consists of vertices $1$ to $8$, and community $B$ consists of vertices $9$ to $11$. When vertex $6$ joins community $B$ due to its strong connection with $11$, it leaves community $A$ internally-disconnected. Thus, in both examples, community $A$ should thus be marked for splitting.}
  \label{fig:about-address-split}
\end{figure}

Now there still remains an issue. At no cost do we want to introduce disconnected communities in the returned community structures. Figure \ref{fig:about-address-split} shows two different cases which may cause a community to be internally-disconnected.\ignore{Such communities should be processed to split apart its disconnected components into separate communities.} First, as shown in the top subfigure, edge deletions within a community may cause it to become internally-disconnected. Therefore, during initialization, any community with intra-community edge deletions should be processed to split apart its disconnected components into separate communities. Second, as shown in the bottom subfigure, when a vertex migrates from its original community to another one --- due to stronger connections --- it can cause the original community to become internally disconnected. Thus, in the local-moving phase of the algorithm, source communities from which vertices migrate should also be processed to separate any disconnected components into distinct communities. Note however that if a community has already been marked for refinement, splitting is not necessary.

In order to address the above mentioned issues, we split all communities that are not being refined, at the end of each pass. This has been show to have better performance than simply splitting communities at the end of all passes \cite{sahu2024approach}. However, we can do better. We can identify a subset of communities to split, and ignore the communities that are guaranteed to have no disconnected communities. Note that a disconnected community may arise only for communities with intra-community edge deletions, and if a vertex migrates from its old community to some other community. Accordingly, we choose to keep track of such communities.

\begin{figure}[!hbt]
  \centering
  \subfigure{
    \label{fig:8020-select-split--all}
    \includegraphics[width=0.98\linewidth]{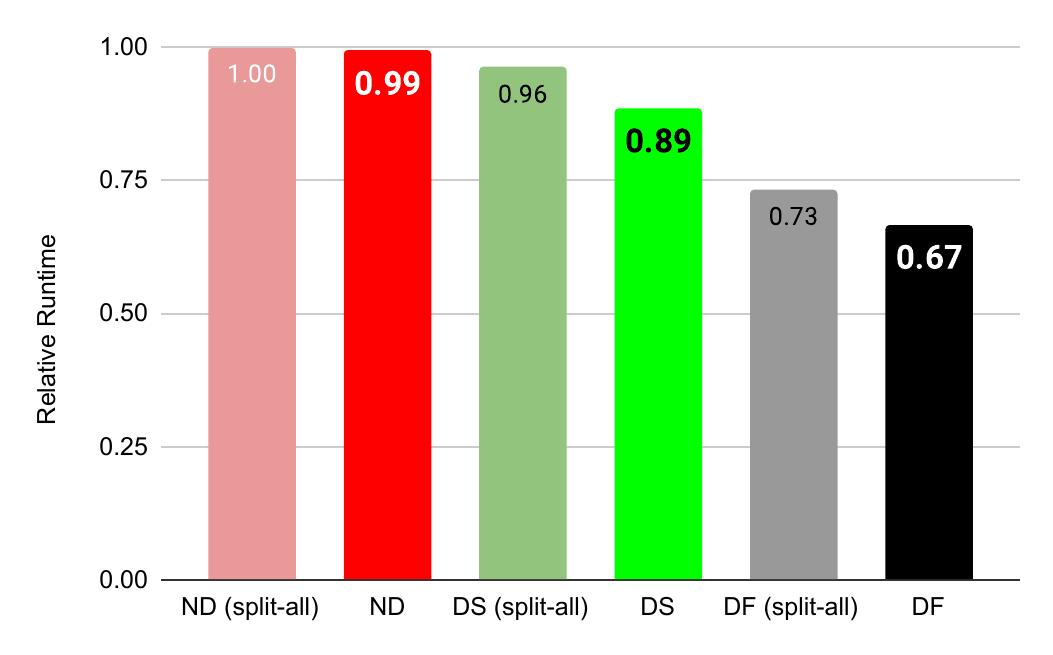}
  } \\[-4ex]
  \caption{Relative Runtime of \textit{Naive-dynamic (ND)}, \textit{Delta-screening (DS)}, and \textit{Dynamic Frontier (DF) Leiden} evaluated under two conditions: (1) splitting all communities that have not been flagged for refinement ("split-all"), and (2) splitting only a subset of communities that were marked during initialization or the local-moving phase of the algorithm.}
  \label{fig:8020-select-split}
\end{figure}

However, our initial observation seemed to indicate that tracking of communities to split has a relatively significant added cost. Accordingly, we conduct an experiment to measure the performance of ND, DS, and DF Leiden under two conditions: splitting all communities that have not been marked to be refined ("split-all"), and splitting only a subset of communities (which have a chance to form internally-disconnected communities), marked during initialization or the local-moving phase of the algorithm. We perform this experiment on large graphs with random batch updates of size ranging from $10^{-7}|E|$ to $0.1|E|$. Figure \ref{fig:8020-select-split} shows the relative runtimes of ND, DS, and DF Leiden. As results show, splitting a subset of communities improves performance, despite added tracking cost.

\subsection{Implementation details}
\label{sec:implementation-details}

We implement an asynchronous version of Leiden, allowing threads to independently process different graph sections, which enhances convergence speed but increases variability in results. Each thread maintains its own hashtable to track delta-modularity during local-moving and refinement phases, as well as total edge weights between super-vertices during aggregation. Optimizations include OpenMP's dynamic loop scheduling, limiting iterations to 20 per pass, a tolerance drop rate of 10 starting at 0.01, vertex pruning, parallel prefix sums, and preallocated Compressed Sparse Row (CSR) structures for super-vertex graphs and community vertices, along with fast, collision-free per-thread hashtables \cite{sahu2024fast}. To ensure high-quality communities, we do not skip the aggregation phase of the\ignore{Leiden} algorithm \cite{sahu2024starting}, even when merging a small number of communities, unlike GVE-Leiden \cite{sahu2024fast}. This results in only a minor increase in runtime across Static, ND, DS, and DF Leiden. We also use a refine-based variation, where community labels of super-vertices are determined by the refinement instead of the local-moving phase, allowing for splitting apart of isolated communities.

The pseudocode for our improved ND, DS, and DF Leiden is given in Sections \ref{sec:our-naive}, \ref{sec:our-delta} and \ref{sec:our-frontier}, respectively. During the first pass of the Leiden algorithm, we process the vertices identified as affected by the DS and DF approaches, initializing each vertex's community membership based on the membership from the previous graph snapshot. In subsequent passes, all super-vertices are designated as affected and processed according to the Leiden algorithm \cite{sahu2024shared}. Similar to the DF Louvain method \cite{sahu2024dflouvain}, we utilize the weighted degrees of vertices $K^{t-1}$, the total edge weights of communities $\Sigma^{t-1}$, and the cumulative change in the total edge weights of communities $\Delta \Sigma^{t-1}$, as auxiliary information to the dynamic algorithm.

\subsection{Time and Space complexity}

The time complexity of ND, DS, and DF Leiden remains $O(L|E^t|)$, similar to Static Leiden, where $L$ is the total number of iterations. However, the local-moving and refinement phases' costs during the first pass are reduced and depend on the batch update's size and nature. The space complexity matches that of Static Leiden, i.e., $O(T|V^t| + |E^t|)$, with $T$ being the number of threads used ($T|V^t|$ accounts for per-thread collision-free hashtables \cite{sahu2024fast}).

\section{Evaluation}
\label{sec:evaluation}
\subsection{Experimental setup}
\label{sec:setup}

\subsubsection{System}
\label{sec:system}

We use a server with an x86-based 64-bit AMD EPYC-7742 processor running at $2.25$ GHz, paired with 512 GB of DDR4 RAM. Each core has a 4 MB L1 cache, a 32 MB L2 cache, and a shared 256 MB L3 cache. The server operates on Ubuntu 20.04.

\subsubsection{Configuration}
\label{sec:configuration}

We use 32-bit unsigned integers for vertex IDs and 32-bit floating-point numbers for edge weights. For floating-point aggregations, we switch to 64-bit floating-point. Affected vertices and communities marked for splitting/refining are represented by 8-bit integer vectors. Key parameters include an iteration tolerance $\tau$ of $10^{-2}$ for the local-moving phase, a refinement tolerance $\tau_{re}$ of $0.6$, a \textit{\small{TOLERANCE\_DECLINE\_FACTOR}} of $10$ for threshold scaling optimization \cite{lu2015parallel}, a \textit{\small{MAX\_ITERATIONS}} of $20$ for the local-moving phase per pass, and a \textit{\small{MAX\_PASSES}} of $10$ \cite{sahu2024fast}. We employ OpenMP's dynamic scheduling with a chunk size of $2048$ for the local-moving, refinement, and aggregation phases. However, for the aggregation phase in Naive-dynamic (ND), Delta-screening (DS), and Dynamic Frontier (DF) Leiden, we use a chunk size of $32$ for proper work balancing \cite{sahu2024starting}. We run all implementations on $64$ threads, unless stated otherwise, and compile using GCC 9.4 with OpenMP 5.0.

\subsubsection{Dataset}
\label{sec:dataset}

We conduct experiments on $12$ large real-world graphs with random batch updates, as listed in Table \ref{tab:dataset-large}, sourced from the SuiteSparse Matrix Collection. These graphs range in size from $3.07$ million to $214$ million vertices, and from $25.4$ million to $3.80$ billion edges. For experiments involving real-world dynamic graphs, we utilize five temporal networks from the Stanford Large Network Dataset Collection \cite{snapnets}, detailed in Table \ref{tab:dataset}. These networks feature vertex counts between $24.8$ thousand and $2.60$ million, with temporal edges ranging from $507$ thousand to $63.4$ million, and static edges from $240$ thousand to $36.2$ million. However, it is worth noting that most temporal graphs in the SNAP repository \cite{snapnets} are relatively small\ignore{(with one exception)}, limiting their applicability for studying our proposed parallel algorithms. In all experiments, we ensure that the edges are undirected and weighted, with a default weight of $1$. Due to the small size of most publicly available real-world weighted graphs, we do not use them in this study, although our parallel algorithms can handle weighted graphs without modification.

Additionally, we exclude SNAP datasets with ground-truth communities because they are non-disjoint, while our focus is on disjoint communities. It is important to note that community detection is not solely about matching ground truth, as this may not accurately represent the network's actual structure and could overlook meaningful community patterns \cite{peel2017ground}.

\subsubsection{Batch generation}
\label{sec:batch-generation}

We take each base graph from Table \ref{tab:dataset-large} and generate random batch updates \cite{com-zarayeneh21} with an $80\% : 20\%$ mix of edge insertions and deletions, each edge weighted at $1$. All updates are undirected; for each insertion $(i, j, w)$, we also include $(j, i, w)$. Vertex pairs are selected uniformly for insertions, while existing edges are uniformly deleted for deletions. No new vertices are added or removed. The batch size, measured as a fraction of the original graph's edges, ranges from $10^{-7}$ to $0.1$, translating to $100$ to $100$ million edges for a billion-edge graph. We conduct five distinct random batch updates for each batch size and report the average results. Do note that dynamic graph algorithms are useful for small batch updates, such as in interactive applications, while static algorithms are generally more efficient for larger batches.

For experiments on real-world dynamic graphs, we initially load $90\%$ of each graph listed in Table \ref{tab:dataset}. As earlier, all edges are assigned a default weight of $1$ and are made undirected by adding their reverse edges. Next, we load $B$ edges for $100$ batch updates, where $B$ represents the batch size as a fraction of the total temporal edges, $|E_T|$, in the graph. Each batch update is ensured to be undirected.

\begin{table}[hbtp]
  \centering
  \caption{List of $12$ graphs retrieved from the SuiteSparse Matrix Collection \cite{suite19} (with directed graphs indicated by $*$). Here, $|V|$ denotes the number of vertices, $|E|$ denotes the number of edges (after making the graph undirected by adding reverse edges), and $|\Gamma|$ denotes the number of communities obtained with \textit{Static Leiden} algorithm \cite{sahu2024fast}.}
  \label{tab:dataset-large}
  \begin{tabular}{|c||c|c|c|}
    \toprule
    \textbf{Graph} &
    \textbf{\textbf{$|V|$}} &
    \textbf{\textbf{$|E|$}} &
    \textbf{\textbf{$|\Gamma|$}} \\
    \midrule
    \multicolumn{4}{|c|}{\textbf{Web Graphs (LAW)}} \\ \hline
    indochina-2004$^*$ & 7.41M & 341M & 2.68K \\ \hline
    arabic-2005$^*$ & 22.7M & 1.21B & 2.92K \\ \hline
    uk-2005$^*$ & 39.5M & 1.73B & 18.2K \\ \hline
    webbase-2001$^*$ & 118M & 1.89B & 2.94M \\ \hline
    it-2004$^*$ & 41.3M & 2.19B & 4.05K \\ \hline
    sk-2005$^*$ & 50.6M & 3.80B & 2.67K \\ \hline
    \multicolumn{4}{|c|}{\textbf{Social Networks (SNAP)}} \\ \hline
    com-LiveJournal & 4.00M & 69.4M & 3.09K \\ \hline
    com-Orkut & 3.07M & 234M & 36 \\ \hline
    \multicolumn{4}{|c|}{\textbf{Road Networks (DIMACS10)}} \\ \hline
    asia\_osm & 12.0M & 25.4M & 2.70K \\ \hline
    europe\_osm & 50.9M & 108M & 6.13K \\ \hline
    \multicolumn{4}{|c|}{\textbf{Protein k-mer Graphs (GenBank)}} \\ \hline
    kmer\_A2a & 171M & 361M & 21.1K \\ \hline
    kmer\_V1r & 214M & 465M & 10.5K \\ \hline
  \bottomrule
  \end{tabular}
\end{table}

\begin{table}[hbtp]
  \centering
  \caption{List of $5$ real-world dynamic graphs obtained from the Stanford Large Network Dataset Collection \cite{snapnets}. Here, $|V|$ denotes the vertex count, $|E_T|$ denotes the total number of temporal edges (including duplicates), and $|E|$ denotes the number of static edges (excluding duplicates).}
  \label{tab:dataset}
  \begin{tabular}{|c||c|c|c|c|}
    \toprule
    \textbf{Graph} &
    \textbf{\textbf{$|V|$}} &
    \textbf{\textbf{$|E_T|$}} &
    \textbf{\textbf{$|E|$}} \\
    \midrule
    sx-mathoverflow & 24.8K & 507K & 240K \\ \hline
    sx-askubuntu & 159K & 964K & 597K \\ \hline
    sx-superuser & 194K & 1.44M & 925K \\ \hline
    wiki-talk-temporal & 1.14M & 7.83M & 3.31M \\ \hline
    sx-stackoverflow & 2.60M & 63.4M & 36.2M \\ \hline
  \bottomrule
  \end{tabular}
\end{table}

\subsubsection{Measurement}
\label{sec:measurement}

We assess the runtime of each method on the entire updated graph, covering all phases of the algorithm. To minimize the impact of noise in our experiments, we adhere to the standard practice of running each experiment multiple times. We assume the total edge weight of the graphs is known and can be tracked with each batch update. As a baseline, we use the most efficient multicore implementation of Static Leiden, GVE-Leiden \cite{sahu2024fast}, which has been shown to outperform even GPU-based approaches. Since modularity maximization is NP-hard and all existing polynomial-time algorithms are heuristic-based, we evaluate the optimality of our dynamic algorithms by comparing their convergence to the modularity score achieved by the static algorithm. Finally, because none of the algorithms analyzed --- Static, ND, DS, and DF Leiden --- produce internally disconnected communities, we omit this detail from our figures.

\begin{figure*}[hbtp]
  \centering
  \subfigure[Overall result]{
    \label{fig:8020-runtime--mean}
    \includegraphics[width=0.38\linewidth]{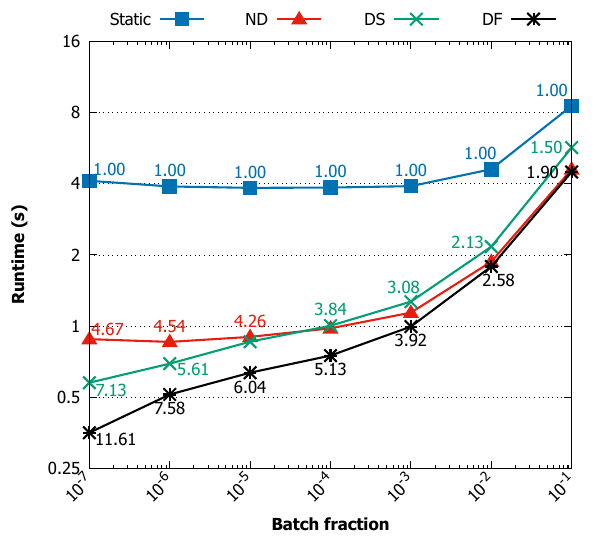}
  }
  \subfigure[Results on each graph]{
    \label{fig:8020-runtime--all}
    \includegraphics[width=0.58\linewidth]{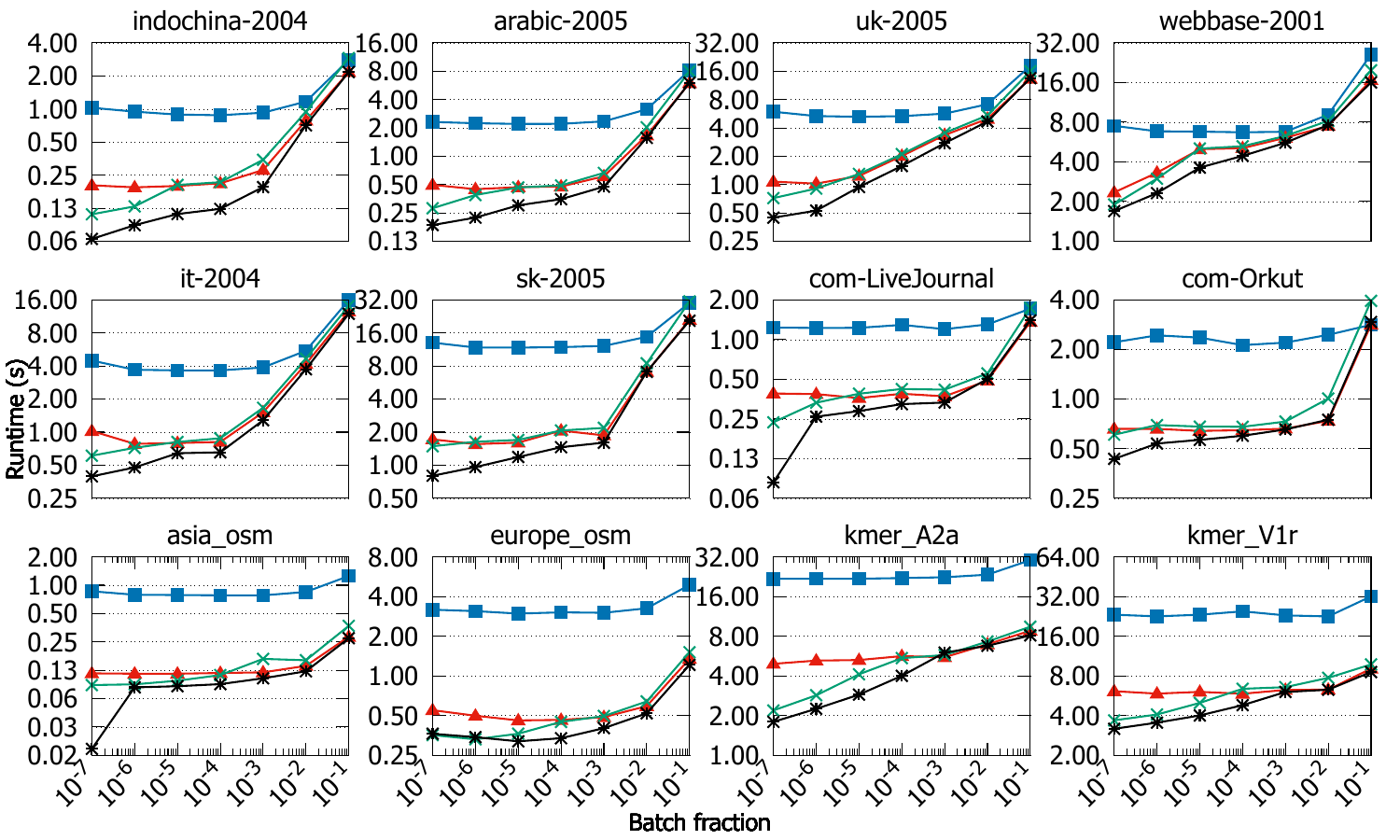}
  } \\[-2ex]
  \caption{Runtime (logarithmic scale) of multicore implementation of our improved \textit{Naive-dynamic (ND)}, \textit{Delta-screening (DS)}, and \textit{Dynamic Frontier (DF) Leiden}, compared to \textit{Static Leiden} \cite{sahu2024fast}, on large\ignore{(static)} graphs with randomly generated batch updates. The size of these batch updates varies from $10^{-7}|E|$ to $0.1|E|$, in multiples of $10$, and consists of $80\%$ edge insertions and $20\%$ edge deletions to mimic realistic dynamic graph modifications. The right subfigure illustrates the runtime for each algorithm across individual graphs in the dataset, while the left subfigure presents the overall runtime using the geometric mean to ensure consistent scaling across graphs.\ignore{Furthermore,} The speedup of each algorithm relative to Static Leiden is shown on the corresponding lines.}
  \label{fig:8020-runtime}
\end{figure*}

\begin{figure*}[hbtp]
  \centering
  \subfigure[Overall result]{
    \label{fig:8020-modularity--mean}
    \includegraphics[width=0.38\linewidth]{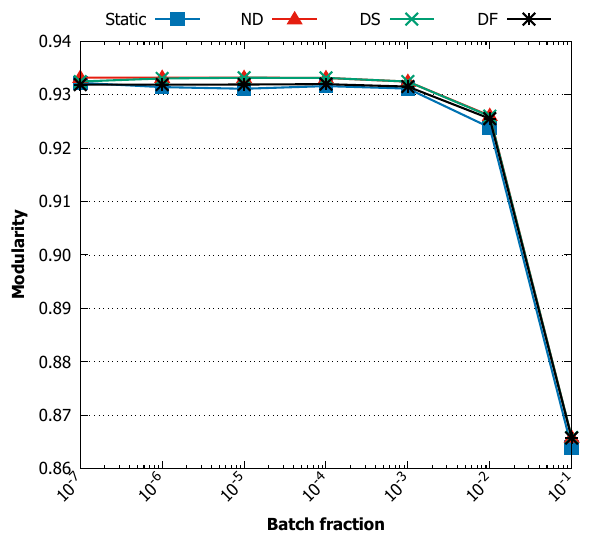}
  }
  \subfigure[Results on each graph]{
    \label{fig:8020-modularity--all}
    \includegraphics[width=0.58\linewidth]{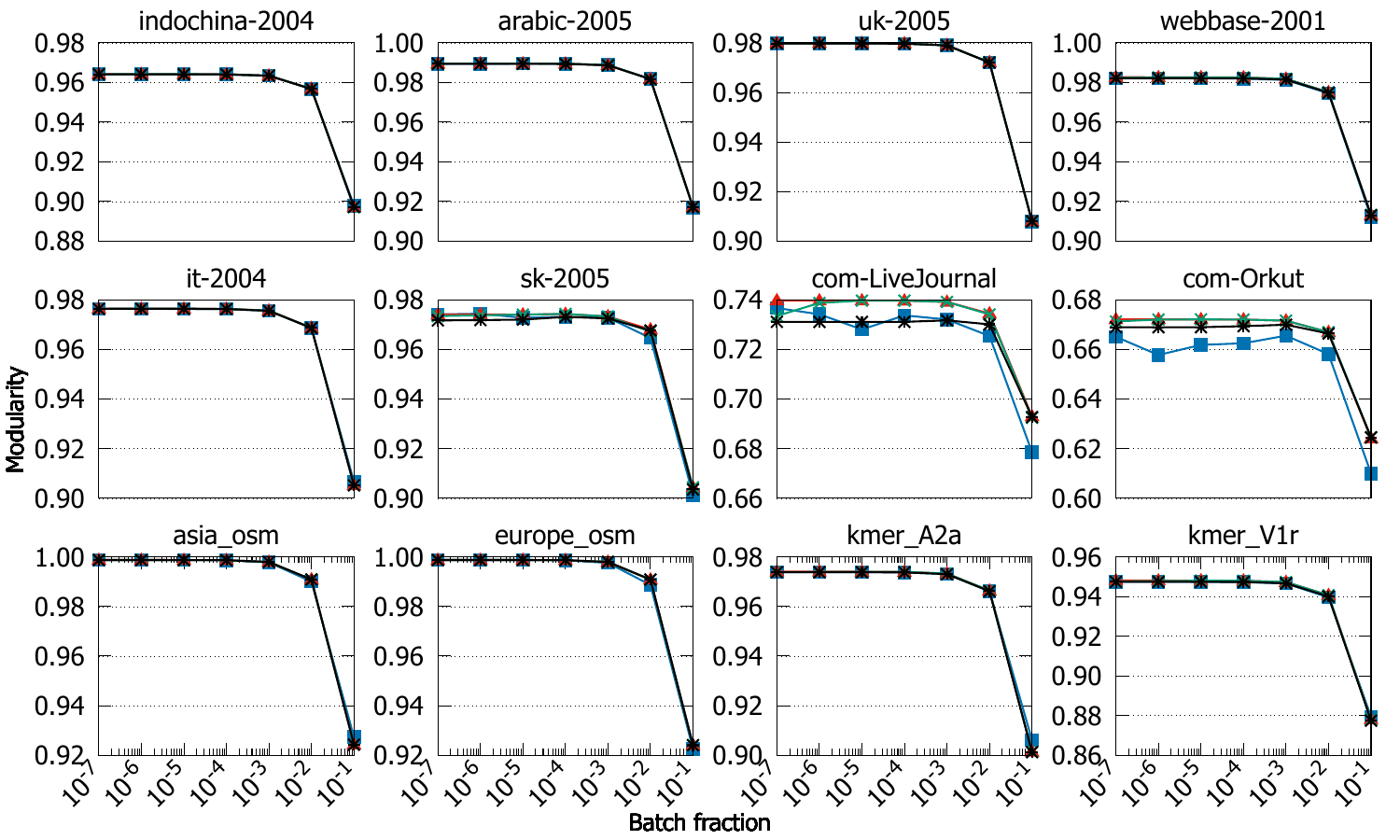}
  } \\[-2ex]
  \caption{Modularity comparison of multicore implementation of our improved \textit{Naive-dynamic (ND)}, \textit{Delta-screening (DS)}, and \textit{Dynamic Frontier (DF) Leiden}, compared to \textit{Static Leiden} \cite{sahu2024fast}, on large\ignore{(static)} graphs that undergo randomly generated batch updates. These updates, which range in size from $10^{-7}|E|$ to $0.1|E|$ in powers of $10$, simulate realistic dynamic graph changes, with $80\%$ edge insertions and $20\%$ edge deletions. Here, the right subfigure presents the modularity for each algorithm on individual graphs, while the left subfigure shows the overall modularity, computed as the arithmetic mean.}
  \label{fig:8020-modularity}
\end{figure*}

\subsection{Performance comparison}
\label{sec:performance-comparison}

We now measure the performance of parallel implementations of our improved ND, DS, and DF Leiden against Static Leiden \cite{sahu2024fast} on large graphs, given in Table \ref{tab:dataset-large}, with randomly generated batch updates. As outlined in Section \ref{sec:batch-generation}, these updates vary in size from $10^{-7}|E|$ to $0.1|E|$, with $80\%$ edge insertions and $20\%$ edge deletions. Each batch update includes reverse edges to maintain an undirected graph structure. Figure \ref{fig:8020-runtime--all} illustrates the execution time of each algorithm for individual graphs, while Figure \ref{fig:8020-runtime--mean} provides a comparison of overall runtimes using the geometric mean for consistent scaling across different graph sizes. Additionally, Figure \ref{fig:8020-modularity--all} presents the modularity results for each graph in the dataset, and Figure \ref{fig:8020-modularity--mean} shows the overall modularity for each algorithm, averaged using the arithmetic mean. Performance comparison on real-world dynamic graphs from Table \ref{tab:dataset} is given in Section \ref{sec:performance-comparison-temporal}.

From Figure \ref{fig:8020-runtime--mean}, we can see that our improved ND, DS, and DF Leiden achieve mean speedups of $3.9\times$, $4.4\times$, and $6.1\times$, respectively, compared to Static Leiden. These speedups are even more pronounced for smaller batch updates, where ND, DS, and DF Leiden provide average speedups of $4.7\times$, $7.1\times$, and $11.6\times$, respectively, for batch updates of size $10^{-7}|E|$. Figure \ref{fig:8020-runtime--all} further illustrates that ND, DS, and DF Leiden deliver significant speedups across all graph types, though the gains are somewhat lower for social networks (with high average degree nodes) and protein k-mer graphs (with low average degree nodes). This may be due to these classes of graphs lacking dense community structures.

In terms of modularity, as shown in Figures \ref{fig:8020-modularity--mean} and \ref{fig:8020-modularity--all}, our improved ND, DS, and DF Leiden achieve communities with modularity scores comparable to those of Static Leiden. However, on social networks, ND, DS, and DF Leiden slightly outperform Static Leiden. This is likely because social networks lack a strong community structure and thus require more iterations to reach a good community assignment. On these graphs, Static Leiden tends to underperform in terms of modularity due to its constraint on the number of iterations per pass, which is designed to prioritize faster runtimes. In contrast, ND, DS, and DF Leiden, which build upon the community membership generated by Static Leiden rather than starting from scratch, are able to refine and improve the modularity.\ignore{As a result, the average modularity from Static Leiden is marginally lower.} Nonetheless, the difference in modularity between Static Leiden and our algorithms is less than $0.002$ on average.

Additionally, note in Figure \ref{fig:8020-runtime} that the runtime of Static Leiden increases with larger batch updates. This is mainly because random batch updates disrupt the existing community structure, requiring Static Leiden to perform more iterations to reach convergence --- it is not solely a consequence of the graph's increased edge count. This disruption also explains the observed decrease in modularity for all algorithms as batch sizes grow, as seen in Figure \ref{fig:8020-modularity}.

\subsection{Affected vertices and Marked communities}

\begin{figure}[hbtp]
  \centering
  \subfigure{
    \label{fig:8020-affected--all}
    \includegraphics[width=0.98\linewidth]{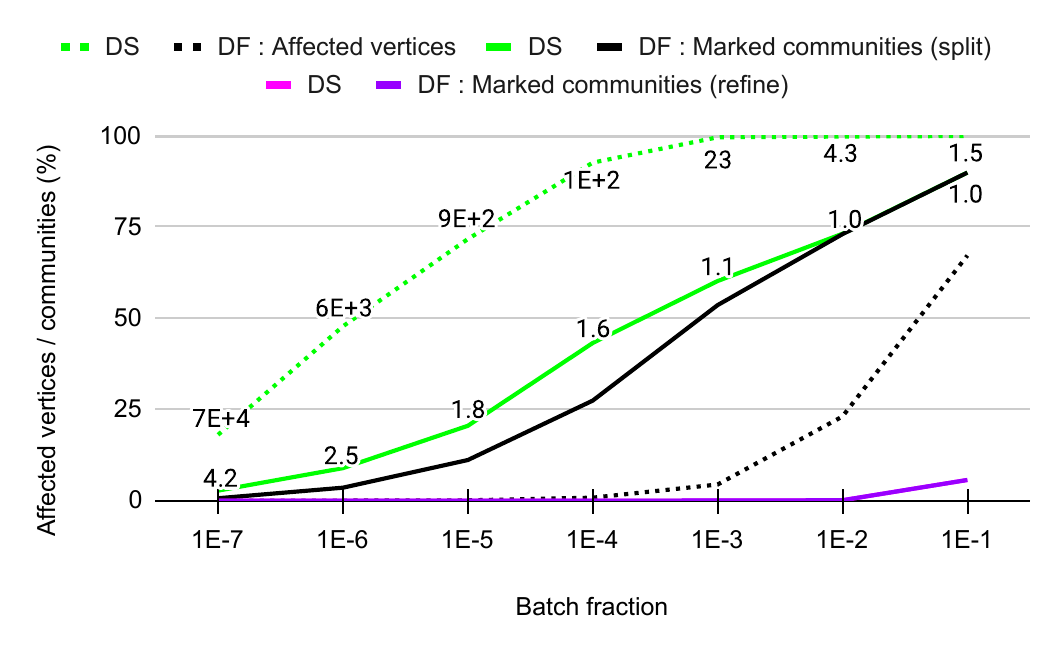}
  } \\[-4ex]
  \caption{Fraction of vertices marked as affected (dotted lines), communities marked to be split (solid lines), and communities marked to be refined (pink, purple lines) with our improved \textit{Delta-screening (DS)} and \textit{Dynamic Frontier (DF) Leiden} on graphs in Table \ref{tab:dataset-large}. The labels indicate the ratio of vertices marked as affected (top) and communities marked to be split (bottom) by DS Leiden to that of DF Leiden. The communities marked to be refined by DS and DF Leiden are nearly identical, and thus not labeled.}
  \label{fig:8020-affected}
\end{figure}

We now analyze the fraction of vertices marked as affected, the fraction of communities marked for splitting, and the fraction of communities marked for refinement by our improved DS and DF Leiden algorithms. This analysis is conducted over the instances in Table \ref{tab:dataset-large}, using batch updates ranging from $10^{-7}|E|$ to $0.1|E|$. It is important to note that we only track affected vertices and marked communities (for splitting and refinement) during the first pass of the Leiden algorithm. These results are presented in Figure \ref{fig:8020-affected}.

From Figure \ref{fig:8020-affected}, we observe that DF Leiden marks significantly fewer vertices as affected compared to DS Leiden. However, the runtime difference between the two algorithms is smaller, as many of the vertices marked by DS Leiden do not change their community labels and converge quickly. Additionally, the number of affected vertices does not fully account for the overall performance differences between the two algorithms. The figure also shows that DF Leiden marks fewer communities for splitting than DS Leiden, though the gap is much smaller than that of affected vertices, and more closely mirrors the performance gap between DS and DF Leiden. Finally, both DS and DF Leiden mark roughly the same small fraction of communities for refinement, thanks to our heuristic that minimizes the number of communities to be refined. This fraction only becomes noticeable at a batch size of $0.1|E|$. As expected, performance deteriorates with an increasing number of affected vertices and marked communities, as shown in Figures \ref{fig:8020-runtime} and \ref{fig:8020-affected}. While a high number of marked communities increases splitting and refinement costs in the first pass, the cost of subsequent passes depends primarily on the nature of the graph.

\begin{figure}[hbtp]
  \centering
  \subfigure{
    \label{fig:8020-scaling--am}
    \includegraphics[width=0.98\linewidth]{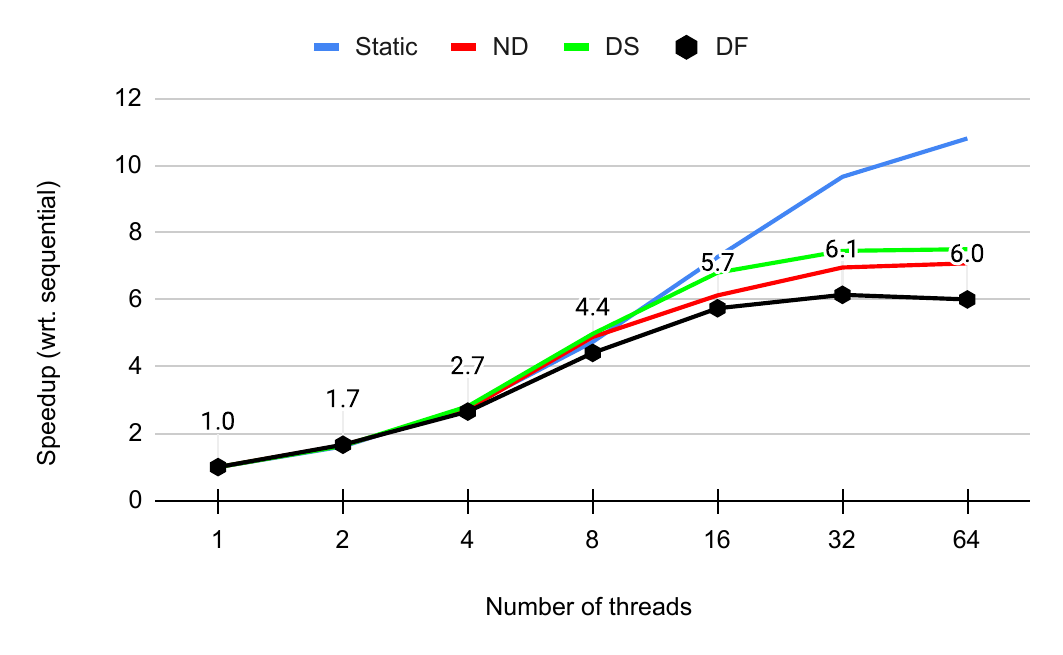}
  } \\[-4ex]
  \caption{Strong scalability of our improved \textit{Naive-dynamic (ND)}, \textit{Delta-screening (DS)}, and \textit{Dynamic Frontier (DF) Leiden}, compared to \textit{Static Leiden}, on batch updates of size $10^{-3} |E|$. The number of threads is doubled from $1$ to $64$ (log-scale).}
  \label{fig:8020-scaling}
\end{figure}

\subsection{Scalability}

Finally, we study the strong-scaling behavior of our improved ND, DS, and DF Leiden, and compare to Static Leiden. For this, we fix the batch size at $10^{-3} |E|$ and vary thread count from $1$ to $64$, measuring the speedup of each algorithm relative to its sequential execution.

As shown in Figure \ref{fig:8020-scaling}, at $32$ threads, ND, DS, and DF Leiden achieve speedups of $7.0\times$, $7.4\times$, and $6.1\times$, respectively, compared to their sequential counterparts. Their speedups increase at average rates of $1.48\times$, $1.49\times$, and $1.44\times$ for every doubling of thread count. However, at $64$ threads, performance is impacted by NUMA effects. Additionally, as thread count increases, the algorithms behave more synchronously --- similar to the Jacobi iterative method --- further reducing performance. The limited scalability is also due to the presence of sequential steps in the algorithms and insufficient work to distribute across threads for the dynamic algorithms.

\section{Conclusion}
\label{sec:conclusion}
This technical report introduced three techniques/heuristics aimed at enhancing the performance and stability of our previously proposed multicore dynamic community detection algorithms based on Leiden \cite{sahu2024starting}. Experiments conducted on a 64-core AMD EPYC-7742 processor show that our improved ND, DS, and DF Leiden algorithms achieve average speedups of $3.9\times$, $4.4\times$, and $6.1\times$, respectively, on large graphs undergoing random batch updates when compared to the Static Leiden algorithm. Additionally, these methods demonstrate a scaling efficiency of $1.4 - 1.5\times$ for each doubling of threads. In the future, we would like to further improve the trackability of communities on evolving graphs.

\begin{acks}
I would like to thank Prof. Kishore Kothapalli and Prof. Dip Sankar Banerjee for their support.
\end{acks}

\bibliographystyle{ACM-Reference-Format}
\bibliography{main}

\clearpage
\appendix
\section{Appendix}
\subsection{Our Parallel Naive-dynamic (ND) Leiden}
\label{sec:our-naive}

Algorithm \ref{alg:naive} presents a multicore implementation of our improved ND Leiden, where vertices are assigned to communities based on the previous graph snapshot. All vertices are processed, regardless of edge deletions or insertions, but the algorithm identifies specific communities that need to be split or refined based on the batch update. It takes as input the current graph snapshot $G^t$, edge deletions $\Delta^{t-}$, insertions $\Delta^{t+}$ in the batch update, previous community memberships $C^{t-1}$, vertex weighted degrees $K^{t-1}$, total community edge weights $\Sigma^{t-1}$, and their changes $\Delta \Sigma^{t-1}$. It returns the updated community memberships $C^t$, vertex weighted degrees $K^t$, and community edge weights $\Sigma^t$ with their changes $\Delta \Sigma^t$.

In the algorithm, two lambda functions, \texttt{isAffected()} (lines \ref{alg:naive--isaff-begin}-\ref{alg:naive--isaff-end}) and \texttt{inAffectedRange()} (lines \ref{alg:naive--isaffrng-begin}-\ref{alg:naive--isaffrng-end}), are defined. These functions indicate that all vertices in the graph $G^t$ should be marked as affected and that these vertices can be incrementally marked as affected, respectively. We then use $K^{t-1}$, $\Sigma^{t-1}$, along with the batch updates $\Delta^{t-}$ and $\Delta^{t+}$, to quickly compute $K^t$ and $\Sigma^t$, which are required for the local-moving phase of the Leiden algorithm (line \ref{alg:naive--auxiliary}). We also utilize the batch update to quickly update $\Delta \Sigma^{t-1}$ to $\Delta \Sigma^t$, which is used to keep track of the cumulative change in total edge weight of each community (line \ref{alg:naive--update-changes}). This is used, along with the batch update, to identify the communities that need to be refined, in $\Delta R$, as well as the communities that need to be split, in $\Delta S$ (line \ref{alg:naive--mark-communities}). These are then employed to run the Leiden algorithm and obtain the updated communities $C^t$ (line \ref{alg:naive--leiden}). Finally, $C^t$ is returned, along with $K^t$, $\Sigma^t$, and $\Delta \Sigma^t$ as the updated auxiliary information (line \ref{alg:naive--return}).

\begin{algorithm}[hbtp]
\caption{Our Parallel \textit{Naive-dynamic (ND)} Leiden \cite{sahu2024starting}.}
\label{alg:naive}
\begin{algorithmic}[1]
\Require{$G^t(V^t, E^t)$: Current/updated input graph}
\Require{$\Delta^{t-}, \Delta^{t+}$: Edge deletions and insertions (batch update)}
\Require{$C^{t-1}, C^t$: Previous, current community of each vertex}
\Require{$K^{t-1}, K^t$: Previous, current weighted-degree of vertices}
\Require{$\Sigma^{t-1}, \Sigma^t$: Previous, current total edge weight of communities}
\Require{$\Delta \Sigma^{t-1}, \Delta \Sigma^t$: Previous, current change in community weights}
\Ensure{$\Delta S, \Delta R$: Communities marked to be split, refined}
\Ensure{$isAffected(i)$: Is vertex $i$ is marked as affected?}
\Ensure{$inAffectedRange(i)$: Can $i$ be incrementally marked?}
\Ensure{$F$: Lambda functions passed to parallel Leiden (Alg. \ref{alg:leiden})}

\Statex

\Function{ndLeiden}{$G^t, \Delta^{t-}, \Delta^{t+}, C^{t-1}, K^{t-1}, \Sigma^{t-1}, \Delta \Sigma^{t-1}$}
  \State $\rhd$ Mark affected vertices
  \Function{isAffected}{$i$} \label{alg:naive--isaff-begin}
    \Return{$1$}
  \EndFunction \label{alg:naive--isaff-end}
  \Function{inAffectedRange}{$i$} \label{alg:naive--isaffrng-begin}
    \Return{$1$}
  \EndFunction \label{alg:naive--isaffrng-end}
  \State $F \gets \{isAffected, inAffectedRange\}$ \label{alg:naive--lambdas}
  \State $\rhd$ Use $K^{t-1}$, $\Sigma^{t-1}$ as auxiliary information (Alg. \ref{alg:update})
  \State $\{K^t, \Sigma^t\} \gets updateWeights(G^t, \Delta^{t-}, \Delta^{t+}, C^{t-1}, K^{t-1}, \Sigma^{t-1})$\label{alg:naive--auxiliary}
  \State $\Delta \Sigma^t \gets updateChanges(G^t, \Delta^{t-}, \Delta^{t+}, C^{t-1}, \Delta \Sigma^{t-1})$ \label{alg:naive--update-changes}
  \State $\rhd$ Mark communities to be split, refined (Alg. \ref{alg:leidenmk})
  \State $\{\Delta S, \Delta R\} \gets markCommunities(G^t, \Delta^{t-}, \Delta^{t+}, C^{t-1}, \Sigma^t, \Delta \Sigma^t)$\label{alg:naive--mark-communities}
  \State $\rhd$ Obtain updated communities (Alg. \ref{alg:leiden})
  \State $C^t \gets leiden(G^t, C^{t-1}, K^t, \Sigma^t, \Delta \Sigma^t, \Delta S, \Delta R, F)$ \label{alg:naive--leiden}
  \Return{$\{C^t, K^t, \Sigma^t, \Delta \Sigma^t\}$} \label{alg:naive--return}
\EndFunction
\end{algorithmic}
\end{algorithm}

\begin{algorithm}[hbtp]
\caption{Our Parallel \textit{Delta-screening (DS)} Leiden \cite{sahu2024starting}.}
\label{alg:delta}
\begin{algorithmic}[1]
\Require{$G^t(V^t, E^t)$: Current/updated input graph}
\Require{$\Delta^{t-}, \Delta^{t+}$: Edge deletions and insertions (batch update)}
\Require{$C^{t-1}, C^t$: Previous, current community of each vertex}
\Require{$K^{t-1}, K^t$: Previous, current weighted-degree of vertices}
\Require{$\Sigma^{t-1}, \Sigma^t$: Previous, current total edge weight of communities}
\Require{$\Delta \Sigma^{t-1}, \Delta \Sigma^t$: Previous, current change in community weights}
\Ensure{$\delta V, \delta E, \delta C$: Is vertex, neighbors, or community affected?}
\Ensure{$H$: Hashtable mapping a community to associated weight}
\Ensure{$\Delta S, \Delta R$: Communities marked to be split, refined}
\Ensure{$isAffected(i)$: Is vertex $i$ is marked as affected?}
\Ensure{$inAffectedRange(i)$: Can $i$ be incrementally marked?}
\Ensure{$F$: Lambda functions passed to parallel Leiden (Alg. \ref{alg:leiden})}

\Statex

\Function{dsLeiden}{$G^t, \Delta^{t-}, \Delta^{t+}, C^{t-1}, K^{t-1}, \Sigma^{t-1}, \Delta \Sigma^{t-1}$}
  \State $H, \delta V, \delta E, \delta C \gets \{\}$ \label{alg:delta--init}
  \State $\rhd$ Mark affected vertices
  \ForAll{$(i, j, w) \in \Delta^{t-}$ \textbf{in parallel}} \label{alg:delta--loopdel-begin}
    \If{$C^{t-1}[i] = C^{t-1}[j]$}
      \State $\delta V[i], \delta E[i], \delta C[C^{t-1}[j]] \gets 1$ \label{alg:delta--loopdelmark}
    \EndIf
  \EndFor \label{alg:delta--loopdel-end}
  \ForAll{unique source vertex $i \in \Delta^{t+}$ \textbf{in parallel}} \label{alg:delta--loopins-begin}
    \State $H \gets \{\}$
    \ForAll{$(i', j, w) \in \Delta^{t+}\ |\ i' = i$} \label{alg:delta--loopinssrc-begin}
      \If{$C^{t-1}[i] \neq C^{t-1}[j]$}
        \State $H[C^{t-1}[j]] \gets H[C^{t-1}[j]] + w$
      \EndIf
    \EndFor \label{alg:delta--loopinssrc-end}
    \State $c^* \gets$ Best community linked to $i$ in $H$ \label{alg:delta--loopinschoose}
    \State $\delta V[i], \delta E[i], \delta C[c^*] \gets 1$ \label{alg:delta--loopinsmark}
  \EndFor \label{alg:delta--loopins-end}
  \ForAll{$i \in V^t$ \textbf{in parallel}} \label{alg:delta--loopaff-begin}
    \If{$\delta E[i]$} \label{alg:delta--loopaffnei-begin}
      \ForAll{$j \in G^t.neighbors(i)$}
        \State $\delta V[j] \gets 1$
      \EndFor
    \EndIf \label{alg:delta--loopaffnei-end}
    \If{$\delta C[C^{t-1}[i]]$} \label{alg:delta--loopaffcom-begin}
      \State $\delta V[i] \gets 1$
    \EndIf \label{alg:delta--loopaffcom-end}
  \EndFor \label{alg:delta--loopaff-end}
  \Function{isAffected}{$i$} \label{alg:delta--isaff-begin}
    \Return{$\delta V[i]$}
  \EndFunction \label{alg:delta--isaff-end}
  \Function{inAffectedRange}{$i$} \label{alg:delta--isaffrng-begin}
    \Return{$\delta V[i]$}
  \EndFunction \label{alg:delta--isaffrng-end}
  \State $F \gets \{isAffected, inAffectedRange\}$ \label{alg:delta--lambdas}
  \State $\rhd$ Use $K^{t-1}$, $\Sigma^{t-1}$ as auxiliary information (Alg. \ref{alg:update})
  \State $\{K^t, \Sigma^t\} \gets updateWeights(G^t, \Delta^{t-}, \Delta^{t+}, C^{t-1}, K^{t-1}, \Sigma^{t-1})$\label{alg:delta--auxiliary}
  \State $\Delta \Sigma^t \gets updateChanges(G^t, \Delta^{t-}, \Delta^{t+}, C^{t-1}, \Delta \Sigma^{t-1})$ \label{alg:delta--update-changes}
  \State $\rhd$ Mark communities to be split, refined (Alg. \ref{alg:leidenmk})
  \State $\{\Delta S, \Delta R\} \gets markCommunities(G^t, \Delta^{t-}, \Delta^{t+}, C^{t-1}, \Sigma^t, \Delta \Sigma^t)$\label{alg:delta--mark-communities}
  \State $\rhd$ Obtain updated communities (Alg. \ref{alg:leiden})
  \State $C^t \gets leiden(G^t, C^{t-1}, K^t, \Sigma^t, \Delta \Sigma^t, \Delta S, \Delta R, F)$ \label{alg:delta--leiden}
  \Return{$\{C^t, K^t, \Sigma^t, \Delta \Sigma^t\}$} \label{alg:delta--return}
\EndFunction
\end{algorithmic}
\end{algorithm}

\subsection{Our Parallel Delta-screening (DS) Leiden}
\label{sec:our-delta}

Algorithm \ref{alg:delta} outlines the pseudocode for multicore implementation of our improved DS Leiden. It uses modularity-based scoring to identify areas of the graph where community membership of vertices is likely to change \cite{com-zarayeneh21}. Input includes the current graph snapshot $G^t$, edge deletions $\Delta^{t-}$ and insertions $\Delta^{t+}$ from the batch update, previous community memberships $C^{t-1}$, weighted vertex degrees $K^{t-1}$, total community edge weights $\Sigma^{t-1}$, and the changes in their total edge weights $\Delta \Sigma^{t-1}$. The output consists of updated community memberships $C^t$, weighted degrees $K^t$, total community edge weights $\Sigma^t$, and changes in the total edge weights $\Delta \Sigma^t$. Before processing, edge deletions $(i, j, w) \in \Delta^{t-}$ and insertions $(i, j, w) \in \Delta^{t+}$ are sorted separately by source vertex ID $i$.

In the algorithm, we begin by initializing a hashtable $H$ that maps communities to their corresponding weights. We also set the affected flags $\delta V$, $\delta E$, and $\delta C$, which indicate whether a vertex, its neighbors, or its community is impacted by the batch update (lines \ref{alg:delta--init}). Edge deletions $\Delta^{t-}$ and insertions $\Delta^{t+}$ are then processed in parallel. For each deletion $(i, j, w) \in \Delta^{t-}$, where vertices $i$ and $j$ are in the same community, we mark the source vertex $i$, its neighbors, and its community as affected (lines \ref{alg:delta--loopdel-begin}-\ref{alg:delta--loopdel-end}). For each unique source vertex $i$ in the insertions $(i, j, w) \in \Delta^{t+}$, if $i$ and $j$ belong to different communities, we identify the community $c^*$ that maximizes the delta-modularity if $i$ were to move to one of its neighboring communities. We then mark $i$, its neighbors, and the community $c^*$ as affected (lines \ref{alg:delta--loopins-begin}-\ref{alg:delta--loopins-end}). Deletions between different communities and insertions within the same community are ignored. Using the affected neighbors $\delta E$ and community flags $\delta C$, we update the affected vertices in $\delta V$ (lines \ref{alg:delta--loopaff-begin}-\ref{alg:delta--loopaff-end}). Next, similar to ND Leiden, we use $K^{t-1}$ and $\Sigma^{t-1}$, along with $\Delta^{t-}$ and $\Delta^{t+}$, to efficiently compute $K^t$ and $\Sigma^t$ (line \ref{alg:delta--auxiliary}). Additionally, we leverage the batch update to quickly adjust $\Delta \Sigma^{t-1}$ to $\Delta \Sigma^t$, which tracks the cumulative change in the total edge weight of each community (line \ref{alg:delta--update-changes}). This information, combined with the batch update, helps identify the communities that require refinement, in $\Delta R$, and those that need to be split, in $\Delta S$ (line \ref{alg:delta--mark-communities}). We then define the necessary lambda functions, \texttt{isAffected()} (lines \ref{alg:delta--isaff-begin}-\ref{alg:delta--isaff-end}) and \texttt{inAffectedRange()} (lines \ref{alg:delta--isaffrng-begin}-\ref{alg:delta--isaffrng-end}), and run the Leiden algorithm to produce updated community assignments $C^t$ (line \ref{alg:delta--leiden}). Finally, we return updated community memberships $C^t$ along with $K^t$, $\Sigma^t$, and $\Delta \Sigma^t$ as updated auxiliary data (line \ref{alg:delta--return}).

\subsection{Our Parallel Dynamic Frontier (DF) Leiden}
\label{sec:our-frontier}

Algorithm \ref{alg:frontier} outlines the pseudocode for our improved parallel DF Leiden implementation. The input includes the updated graph snapshot $G^t$, edge deletions $\Delta^{t-}$ and insertions $\Delta^{t+}$ in the batch update, the previous community assignments $C^{t-1}$ for each vertex, weighted degrees $K^{t-1}$ of vertices, total edge weights $\Sigma^{t-1}$ of communities, and the change in total edge weights $\Delta \Sigma^{t-1}$ of communities. The algorithm produces the updated community memberships $C^t$, weighted degrees $K^t$ of vertices, total edge weights $\Sigma^t$, and the change in total edge weights $\Delta \Sigma^t$ of communities.

\begin{algorithm}[hbtp]
\caption{Our Parallel \textit{Dynamic Frontier (DF)} Leiden \cite{sahu2024dflouvain}.}
\label{alg:frontier}
\begin{algorithmic}[1]
\Require{$G^t(V^t, E^t)$: Current/updated input graph}
\Require{$\Delta^{t-}, \Delta^{t+}$: Edge deletions and insertions (batch update)}
\Require{$C^{t-1}, C^t$: Previous, current community of each vertex}
\Require{$K^{t-1}, K^t$: Previous, current weighted-degree of vertices}
\Require{$\Sigma^{t-1}, \Sigma^t$: Previous, current total edge weight of communities}
\Require{$\Delta \Sigma^{t-1}, \Delta \Sigma^t$: Previous, current change in community weights}
\Ensure{$\delta V$: Flag vector indicating if each vertex is affected}
\Ensure{$\Delta S, \Delta R$: Communities marked to be split, refined}
\Ensure{$isAffected(i)$: Is vertex $i$ is marked as affected?}
\Ensure{$inAffectedRange(i)$: Can $i$ be incrementally marked?}
\Ensure{$onChange(i)$: What happens if $i$ changes its community?}
\Ensure{$F$: Lambda functions passed to parallel Leiden (Alg. \ref{alg:leiden})}

\Statex

\Function{dfLeiden}{$G^t, \Delta^{t-}, \Delta^{t+}, C^{t-1}, K^{t-1}, \Sigma^{t-1}, \Delta \Sigma^{t-1}$}
  \State $\rhd$ Mark initial affected vertices
  \ForAll{$(i, j) \in \Delta^{t-}$ \textbf{in parallel}} \label{alg:frontier--loopdel-begin}
    \If{$C^{t-1}[i] = C^{t-1}[j]$} $\delta V[i] \gets 1$
    \EndIf
  \EndFor \label{alg:frontier--loopdel-end}
  \ForAll{$(i, j, w) \in \Delta^{t+}$ \textbf{in parallel}} \label{alg:frontier--loopins-begin}
    \If{$C^{t-1}[i] \neq C^{t-1}[j]$} $\delta V[i] \gets 1$
    \EndIf
  \EndFor \label{alg:frontier--loopins-end}
  \Function{isAffected}{$i$} \label{alg:frontier--isaff-begin}
    \Return{$\delta V[i]$}
  \EndFunction \label{alg:frontier--isaff-end}
  \Function{inAffectedRange}{$i$} \label{alg:frontier--isaffrng-begin}
    \Return{$1$}
  \EndFunction \label{alg:frontier--isaffrng-end}
  \Function{onChange}{$i$} \label{alg:frontier--onchg-begin}
    \ForAll{$j \in G^t.neighbors(i)$} $\delta V[j] \gets 1$
    \EndFor
  \EndFunction \label{alg:frontier--onchg-end}
  \State $F \gets \{isAffected, inAffectedRange, onChange\}$ \label{alg:frontier--lambdas}
  \State $\rhd$ Use $K^{t-1}$, $\Sigma^{t-1}$ as auxiliary information (Alg. \ref{alg:update})
  \State $\{K^t, \Sigma^t\} \gets updateWeights(G^t, \Delta^{t-}, \Delta^{t+}, C^{t-1}, K^{t-1}, \Sigma^{t-1})$\label{alg:frontier--auxiliary}
  \State $\Delta \Sigma^t \gets updateChanges(G^t, \Delta^{t-}, \Delta^{t+}, C^{t-1}, \Delta \Sigma^{t-1})$\label{alg:frontier--update-changes}
  \State $\rhd$ Mark communities to be split, refined (Alg. \ref{alg:leidenmk})
  \State $\{\Delta S, \Delta R\} \gets markCommunities(G^t, \Delta^{t-}, \Delta^{t+}, C^{t-1}, \Sigma^t, \Delta \Sigma^t)$\label{alg:frontier--mark-communities}
  \State $\rhd$ Obtain updated communities (Alg. \ref{alg:leiden})
  \State $C^t \gets leiden(G^t, C^{t-1}, K^t, \Sigma^t, \Delta \Sigma^t, \Delta S, \Delta R, F)$ \label{alg:frontier--leiden}
  \Return{$\{C^t, K^t, \Sigma^t, \Delta \Sigma^t\}$} \label{alg:frontier--return}
\EndFunction
\end{algorithmic}
\end{algorithm}

In the algorithm, we begin by identifying a set of affected vertices whose community memberships might change due to batch updates. These vertices are marked in the flag vector $\delta V$. Specifically, we mark the endpoints of deleted edges $\Delta^{t-}$ that belong to the same community (lines \ref{alg:frontier--loopdel-begin}-\ref{alg:frontier--loopdel-end}), as well as the endpoints of inserted edges $\Delta^{t+}$ that connect vertices from different communities (lines \ref{alg:frontier--loopins-begin}-\ref{alg:frontier--loopins-end}). Next, three lambda functions are defined for the Leiden algorithm: \texttt{isAffected()} (lines \ref{alg:frontier--isaff-begin}-\ref{alg:frontier--isaff-end}), \texttt{inAffectedRange()} (lines \ref{alg:frontier--isaffrng-begin}-\ref{alg:frontier--isaffrng-end}), and \texttt{onChange()} (lines \ref{alg:frontier--onchg-begin}-\ref{alg:frontier--onchg-end}). These functions define how vertices are marked as affected: initially through the batch update, incrementally during graph traversal, and when a vertex changes its community, respectively. Importantly, the set of affected vertices grows automatically due to the vertex pruning optimization used in our Parallel Leiden algorithm (Algorithm \ref{alg:leiden}). Here, \texttt{onChange()} simulates the effect of a plain DF approach without pruning. Our approach leverages the previous state of the graph, $K^{t-1}$ and $\Sigma^{t-1}$, alongside the batch updates $\Delta^{t-}$ and $\Delta^{t+}$, to efficiently compute the new states $K^t$ and $\Sigma^t$, which are essential for the local-moving phase of the Leiden algorithm (line \ref{alg:frontier--auxiliary}). We also use the batch update to quickly update $\Delta \Sigma^{t-1}$ to $\Delta \Sigma^t$, which tracks the cumulative change in each community's total edge weight (line \ref{alg:frontier--update-changes}). This information, together with the batch update, helps us identify the communities that require refinement in $\Delta R$ and those that need to be split in $\Delta S$ (line \ref{alg:frontier--mark-communities}). The lambda functions are then employed to execute the Leiden algorithm and generate the updated community assignments, $C^t$ (line \ref{alg:frontier--leiden}). Finally, we return $C^t$, along with $K^t$, $\Sigma^t$, and $\Delta \Sigma^t$ as the updated auxiliary information (line \ref{alg:frontier--return}).

\subsection{Our Dynamic-supporting Parallel Leiden}
\label{sec:our-leiden}

The main step of our Dynamic-supporting Parallel Leiden is presented in Algorithm \ref{alg:leiden}. Unlike our Static Leiden implementation \cite{sahu2024fast}, this algorithm takes into account not only the current graph snapshot $G^t$ but also the previous community memberships $C^{t-1}$ of each vertex, the updated weighted degree $K^t$ of each vertex, the updated total edge weight $\Sigma^t$ of each community, and the change in total edge weight $\Delta \Sigma^t$ of each community. Additionally, it accepts flag vectors $\Delta S$ and $\Delta R$, which indicate whether a community should be split or refined, along with a set of lambda functions $F$ that determine if a vertex is affected or can be progressively identified as affected within the affected range. Finally, updated community memberships of vertices, denoted as $C^t = C$, are returned.

\begin{algorithm}[hbtp]
\caption{Our Dynamic-supporting Parallel Leiden \cite{sahu2024starting}.}
\label{alg:leiden}
\begin{algorithmic}[1]
\Require{$G^t(V^t, E^t)$: Current input graph}
\Require{$C^{t-1}$: Previous community of each vertex}
\Require{$K^t$: Current weighted-degree of each vertex}
\Require{$\Sigma^t$: Current total edge weight of each community}
\Require{$\Delta \Sigma^t$: Change in total edge weight of each community}
\Require{$\Delta S, \Delta R$: Communities marked to be split, refined}
\Require{$F$: Lambda functions passed to parallel Leiden}
\Ensure{$G'(V', E')$: Current/super-vertex graph}
\Ensure{$C, C'$: Current community of each vertex in $G^t$, $G'$}
\Ensure{$K, K'$: Current weighted-degree of each vertex in $G^t$, $G'$}
\Ensure{$\Sigma, \Sigma'$: Current total edge weight of each community in $G^t$, $G'$}
\Ensure{$C'_B$: Community bound of each vertex}
\Ensure{$\tau$: Iteration tolerance}

\Statex

\Function{leiden}{$G^t, C^{t-1}, K^t, \Sigma^t, \Delta \Sigma^t, \Delta S, \Delta R, F$} \label{alg:leiden--begin}
  \State $\rhd$ Mark affected vertices as unprocessed
  \ForAll{$i \in V^t$} \label{alg:leiden--mark-begin}
    \If{$F.isAffected(i)$} Mark $i$ as unprocessed
    \EndIf
  \EndFor \label{alg:leiden--mark-end}
  \State $\rhd$ Initialization phase
  \State Vertex membership: $C \gets [0 .. |V^t|)$ \label{alg:leiden--init-begin}
  \State $G' \gets G^t$ \textbf{;} $C' \gets C^{t-1}$ \textbf{;} $K' \gets K^t$ \textbf{;} $\Sigma' \gets \Sigma^t$
  \State $\rhd$ Local-moving and aggregation phases
  \ForAll{$l_p \in [0 .. \text{\small{MAX\_PASSES}})$} \label{alg:leiden--passes-begin}
    \State $l_i \gets leidenMove(G', C', K', \Sigma', \Delta S, F)$ \Comment{Alg. \ref{alg:leidenlm}} \label{alg:leiden--local-move}
    \State $leidenSubsetRenumber(G', C', \Sigma', \Delta S, \Delta R)$ \Comment{Alg. \ref{alg:leidensr}} \label{alg:leiden-subset-renumber}
    \State $C' \gets leidenSplit(G', C', \Delta S, \Delta R)$ \Comment{Alg. \ref{alg:leidensp}} \label{alg:leiden--split-communities}
    \State $C'_B \gets C'$ \Comment{Community bounds for refinement phase}
    \State $leidenRefine(G', C'_B, C', K', \Sigma', \Delta R, \tau)$ \Comment{Alg. \ref{alg:leidenre}} \label{alg:leiden--refine}
    \If{$l_i \le 1$ \textbf{and} no communities refined} \textbf{break}
    \EndIf
    \If{$l_i \le 1$ \textbf{and} \textbf{not} first pass} \textbf{break} \Comment{Done?} \label{alg:leiden--globally-converged}
    \EndIf
    \State $C' \gets$ Renumber communities in $C'$ \label{alg:leiden--renumber}
    \State $C \gets$ Lookup dendrogram using $C$ to $C'$ \label{alg:leiden--lookup}
    \State $G' \gets leidenAggregate(G', C')$ \Comment{Alg. \ref{alg:leidenag}} \label{alg:leiden--aggregate}
    \State $\Sigma' \gets K' \gets$ Weight of each vertex in $G'$ \label{alg:leiden--reset-weights}
    \State Mark all vertices in $G'$ as unprocessed \label{alg:leiden--reset-affected}
    \State $C' \gets [0 .. |V'|)$ \Comment{Use refine-based membership} \label{alg:leiden--useparent}
    \State $\Delta R \gets \{1\ \forall\ V'\}$ \Comment{Refine all communities next pass} \label{alg:leiden--all-communities-changed}
    \State $\tau \gets \tau / \text{\small{TOLERANCE\_DROP}}$ \Comment{Threshold scaling} \label{alg:leiden--threshold-scaling}
  \EndFor \label{alg:leiden--passes-end}
  \State $C \gets$ Lookup dendrogram using $C$ to $C'$ \label{alg:leiden--lookup-last}
  \State $C \gets leidenTrack(G^t, C, C^{t-1}, K^t)$ \Comment{Alg. \ref{alg:leidentr}} \label{alg:leiden--track-communities}
  \Return{$C$} \label{alg:leiden--return}
\EndFunction \label{alg:leiden--end}
\end{algorithmic}
\end{algorithm}

\begin{algorithm}[hbtp]
\caption{Mark communities to be split, refined \cite{sahu2024fast}.}
\label{alg:leidenmk}
\begin{algorithmic}[1]
\Require{$G^t(V^t, E^t)$: Current input graph}
\Require{$\Delta^{t-}, \Delta^{t+}$: Edge deletions and insertions (batch update)}
\Require{$C^{t-1}$: Previous community of each vertex}
\Require{$\Sigma^t$: Current total edge weight of each community}
\Require{$\Delta \Sigma^t$: Change in total edge weight of communities}
\Ensure{$\Delta R, \Delta S$: Communities marked to be refined, split}
\Ensure{$\tau_{re}$: Refinement tolerance}

\Statex

\Function{markCommunities}{$G^t, \Delta^{t-}, \Delta^{t+}, C^{t-1}, \Sigma^t, \Delta \Sigma^t$}
  \State $\Delta R \gets \Delta S \gets \{\}$
  \If{is dynamic alg.}
    \ForAll{$(i, j) \in \Delta^{t-}$ \textbf{in parallel}}
      \State $c \gets C^{t-1}[i]$
      \State $d \gets C^{t-1}[j]$
      \If{$c = d$} $\Delta S[i] \gets 1$
      \EndIf
    \EndFor
    \ForAll{$(i, j) \in \Delta^{t-} \cup \Delta^{t+}$ \textbf{in parallel}}
      \State $c \gets C^{t-1}[i]$
      \State $d \gets C^{t-1}[j]$
      \If{$c = d$ \textbf{and} $\Delta \Sigma^t[c] / \Sigma^t[c] > \tau_{re}$} $\Delta R[i] \gets 1$
      \EndIf
    \EndFor
  \Else\ $\Delta R \gets \{1\ \forall\ V^t\}$
  \EndIf
  \Return{$\{\Delta S, \Delta R\}$}
\EndFunction
\end{algorithmic}
\end{algorithm}

The algorithm begins by marking the affected vertices as unprocessed (lines \ref{alg:leiden--mark-begin}-\ref{alg:leiden--mark-end}), and initializing the community membership $C$ for each vertex in $G^t$, among others. Once initialization is complete, a series of passes are conducted, limited by $MAX\_PASSES$. Each pass consists of local-moving, splitting, refinement, and aggregation phases (lines \ref{alg:leiden--passes-begin}-\ref{alg:leiden--passes-end}). In each pass, the Leiden algorithm’s local-moving phase (Algorithm \ref{alg:leidenlm}) is executed on the affected (unprocessed) vertices to optimize community assignments while identifying communities that may require splitting, recorded in $\Delta S$ (line \ref{alg:leiden--local-move}). Communities identified in $\Delta S$ that have not been marked for refinement are then split to prevent the emergence of disconnected communities (line \ref{alg:leiden--split-communities}). This process involves renumbering the communities by the ID of a vertex within each community using the \texttt{leidenSubsetRename()} function (line \ref{alg:leiden-subset-renumber}). Renumbering is crucial since only a subset of communities will be split or refined, and using the ID of a contained vertex helps avoid two disconnected communities from sharing the same ID. The total edge weights $\Sigma'$ of communities and the flags for communities to be split and refined in $\Delta S$ and $\Delta R$ are updated accordingly. Next, the identified communities for refinement in $\Delta R$ undergo the refinement phase, executed by \texttt{leidenRefine()}. This phase optimizes the community assignments for each vertex within their respective community boundaries $C'_B$ obtained from the local-moving phase (line \ref{alg:leiden--refine}). If the local-moving phase converged after a single iteration during the first pass or if no communities were refined, convergence is achieved, and the community detection process halts (line \ref{alg:leiden--globally-converged}). Conversely, if convergence is not reached, the aggregation phase is initiated. This phase includes renumbering the communities, updating the top-level community memberships $C$ using a dendrogram lookup, performing the aggregation process (Algorithm \ref{alg:leidenag}), and updating the weighted degrees of vertices $K'$ and total edge weights of communities $\Sigma'$ in the super-vertex graph. To prepare for the next pass, all vertices in the graph are marked as unprocessed, community memberships are initialized based on the refinement phase, all communities are marked for refinement for the upcoming pass, and the convergence threshold $\tau$ is adjusted by scaling (line \ref{alg:leiden--threshold-scaling}). The next pass then begins (line \ref{alg:leiden--passes-begin}). After completing all passes, the top-level community memberships $C$ for each vertex in $G^t$ are updated through dendrogram lookup (lines \ref{alg:leiden--lookup-last}-\ref{alg:leiden--return}). This information is subsequently fed into the community tracking algorithm, \texttt{leidenTrack()}, which facilitates tracking communities even after they have been renumbered due to splits or refinements. Finally, the updated community memberships are returned.

\subsubsection{Local-moving phase of our Parallel Leiden}

The pseudocode for the local-moving phase of our Parallel Leiden algorithm is outlined in Algorithm \ref{alg:leidenlm}. This phase iteratively moves vertices between communities in order to maximize modularity. Here, the \texttt{leidenMove()} function takes the current graph $G'$, community membership $C'$, total edge weight $K'$ of each vertex, total edge weight $\Sigma'$ of each community, a flag vector $\Delta S$ for communities marked to be split, and a set of lambda functions as inputs, and ultimately returns the number of iterations executed $l_i$.

\begin{algorithm}[hbtp]
\caption{Local-moving phase of our Parallel Leiden \cite{sahu2024fast}.}
\label{alg:leidenlm}
\begin{algorithmic}[1]
\Require{$G'(V', E')$: Input/super-vertex graph}
\Require{$C'$: Community membership of each vertex}
\Require{$K'$: Total edge weight of each vertex}
\Require{$\Sigma'$: Total edge weight of each community}
\Require{$\Delta S$: Communities marked to be split}
\Require{$F$: Lambda functions passed to parallel Leiden}
\Ensure{$H_t$: Collision-free per-thread hashtable}
\Ensure{$l_i$: Number of iterations performed}
\Ensure{$\tau$: Per iteration tolerance}

\Statex

\Function{leidenMove}{$G', C', K', \Sigma', \Delta S', F$} \label{alg:leidenlm--move-begin}
  \ForAll{$l_i \in [0 .. \text{\small{MAX\_ITERATIONS}})$} \label{alg:leidenlm--iterations-begin}
    \State Total delta-modularity per iteration: $\Delta Q \gets 0$ \label{alg:leidenlm--init-deltaq}
    \ForAll{unprocessed $i \in V'$ \textbf{in parallel}} \label{alg:leidenlm--loop-vertices-begin}
      \State Mark $i$ as processed (prune) \label{alg:leidenlm--prune}
      \If{\textbf{not} $F.inAffectedRange(i)$} \textbf{continue} \label{alg:leidenlm--affrng}
      \EndIf
      \State $H_t \gets scanCommunities(\{\}, G', C', i, false)$ \label{alg:leidenlm--scan}
      \State $\rhd$ Use $H_t, K', \Sigma'$ to choose best community
      \State $c^* \gets$ Best community linked to $i$ in $G'$ \label{alg:leidenlm--best-community-begin}
      \State $\delta Q^* \gets$ Delta-modularity of moving $i$ to $c^*$ \label{alg:leidenlm--best-community-end}
      \If{$c^* = C'[i]$} \textbf{continue} \label{alg:leidenlm--best-community-same}
      \EndIf
      \State $\Sigma'[C'[i]]\ \text{-=}\ K'[i]$ \textbf{;} $\Sigma'[c^*]\ \text{+=}\ K'[i]$ \textbf{atomic} \label{alg:leidenlm--perform-move-begin}
      \State $C'[i] \gets c^*$ \textbf{;} $\Delta Q \gets \Delta Q + \delta Q^*$ \label{alg:leidenlm--perform-move-end}
      \State Mark neighbors of $i$ as unprocessed \label{alg:leidenlm--remark}
      \If{is dynamic alg.} $\Delta S[c] \gets 1$ \label{alg:leidenlm--mark-split-communities}
      \EndIf
    \EndFor \label{alg:leidenlm--loop-vertices-end}
    \If{$\Delta Q \le \tau$} \textbf{break} \Comment{Locally converged?} \label{alg:leidenlm--locally-converged}
    \EndIf
  \EndFor \label{alg:leidenlm--iterations-end}
  \Return{$l_i$} \label{alg:leidenlm--return}
\EndFunction \label{alg:leidenlm--move-end}

\Statex

\Function{scanCommunities}{$H_t, G', C', i, self$}
  \ForAll{$(j, w) \in G'.edges(i)$}
    \If{\textbf{not} $self$ \textbf{and} $i = j$} \textbf{continue}
    \EndIf
    \State $H_t[C'[j]] \gets H_t[C'[j]] + w$
  \EndFor
  \Return{$H_t$}
\EndFunction
\end{algorithmic}
\end{algorithm}


Lines \ref{alg:leidenlm--iterations-begin}-\ref{alg:leidenlm--iterations-end} outline the core loop of the local-moving phase. In line \ref{alg:leidenlm--init-deltaq}, we initialize the total delta-modularity for each iteration, denoted as $\Delta Q$. Then, in lines \ref{alg:leidenlm--loop-vertices-begin}-\ref{alg:leidenlm--loop-vertices-end}, we iterate concurrently over unprocessed vertices. For each vertex $i$, we execute vertex pruning by marking it as processed (line \ref{alg:leidenlm--prune}). We then check whether $i$ is within the affected range (i.e., it can be incrementally marked as affected). If it is not, we move on to the next vertex (line \ref{alg:leidenlm--affrng}). For each vertex $i$ that is not skipped, we scan the communities connected to it (line \ref{alg:leidenlm--scan}), excluding $i$ itself, to identify the optimal community $c^*$ for moving $i$ to (line \ref{alg:leidenlm--best-community-begin}). We then compute the delta-modularity for this potential move (line \ref{alg:leidenlm--best-community-end}), update the community membership of $i$ (lines \ref{alg:leidenlm--perform-move-begin}-\ref{alg:leidenlm--perform-move-end}), and mark its neighbors as unprocessed (line \ref{alg:leidenlm--remark}) if a better community is found.\ignore{Notably, this practice of marking the neighbors of $i$ as unprocessed --- part of the vertex pruning optimization --- also aligns with the DF Leiden algorithm, which marks neighbors as affected when a vertex changes communities. This allows for an incremental expansion of the affected vertex set without additional coding.} Additionally, for ND, DS, and DF Leiden, we designate the source community $c$ of the migrating vertex $i$ as a candidate community to be split in $\Delta S$. In the case of Static Leiden, all communities are refined, and thus do not require splitting. In line \ref{alg:leidenlm--locally-converged}, we check for convergence of the local-moving phase; if convergence is achieved or if the maximum number of iterations\ignore{($MAX\_ITERATIONS$)} is reached, the loop terminates. Finally, in line \ref{alg:leidenlm--return}, we return the total number of iterations performed, denoted as $l_i$.

\subsubsection{Splitting phase of our Parallel Leiden}
\label{sec:splitbfs}

Next, we will outline the pseudocode for the splitting phase, as detailed in Algorithm \ref{alg:leidensp}. This phase employs a parallel Breadth First Search (BFS) technique to partition disconnected communities. The function \texttt{leidenSplit()} takes as input the graph $G'$ for the current pass, the community membership $C'$ for each vertex, and the flag vectors $\Delta S$ and $\Delta R$, which indicate whether each community is marked for splitting or refining. It returns the updated community membership $C''$ for each vertex, where all the disconnected communities have been split into separate communities.

\begin{algorithm}[hbtp]
\caption{Split disconnected communities using BFS \cite{sahu2024approach}.}
\label{alg:leidensp}
\begin{algorithmic}[1]
\Require{$G'(V', E')$: Input/super-vertex graph}
\Require{$C'$: Initial community membership of each vertex}
\Require{$\Delta S, \Delta R$: Communities marked to be split, refined}
\Ensure{$C''$: Updated community membership of each vertex}
\Ensure{$f_{if}$: Perform BFS to vertex $j$ if condition satisfied}
\Ensure{$f_{do}$: Perform operation after each vertex is visited}
\Ensure{$busy$: Is a community being processed by a thread?}
\Ensure{$vis$: Visited flag for each vertex}

\Statex

\Function{leidenSplit}{$G', C', \Delta S, \Delta R$} \label{alg:leidensp--begin}
  \State $C'' \gets vis \gets busy \gets \{\}$ \label{alg:leidensp--init-begin}
  \ForAll{$i \in V'$ \textbf{in parallel}}
    \State $c' \gets C'[i]$
    \If{$\Delta S[c] = 1$ \textbf{and} $\Delta R[c] = 0$} $C''[i] \gets i$
    \Else\ $C''[i] \gets C'[i]$ \textbf{;} $vis[i] \gets 1$
    \EndIf
  \EndFor \label{alg:leidensp--init-end}
  \ForAll{\textbf{threads}} \label{alg:leidensp--threads-begin}
    \ForAll{$i \in V'$} \label{alg:leidensp--loop-begin}
      \State $c' \gets C'[i]$ \textbf{;} $c'' \gets C''[i]$ \label{alg:leidensp--loopinit}
      \If{$vis[i]$ \textbf{or} $busy[c']$} \textbf{continue} \label{alg:leidensp--work}
      \EndIf
      \If{$atomicCAS(busy[c'], 0, 1)$ \textbf{fails}} \textbf{continue} \label{alg:leidensp--atomiccas}
      \EndIf
      \State $f_{if} \gets (j) \implies C'[j] = C'[j]$
      \State $f_{do} \gets (j) \implies C''[j] \gets c''$
      \State $bfsVisitForEach(vis, G', i, f_{if}, f_{do})$ \label{alg:leidensp--bfs}
      \State $busy[c'] \gets 0$
    \EndFor \label{alg:leidensp--loop-end}
  \EndFor \label{alg:leidensp--threads-end}
  \Return{$C''$} \label{alg:leidensp--return}
\EndFunction \label{alg:leidensp--end}
\end{algorithmic}
\end{algorithm}

At the outset, lines \ref{alg:leidensp--init-begin}-\ref{alg:leidensp--init-end} initialize the flag vector $vis$, which indicates whether vertices have been visited, the flag vector $busy$, which shows if a community is currently under processing by a thread, and the labels $C''$ for each vertex, which are set to their respective vertex IDs. Following this, each thread processes every vertex $i$ in the graph $G'$ concurrently (lines \ref{alg:leidensp--loop-begin}-\ref{alg:leidensp--loop-end}). If vertex $i$ has already been visited, or if the community $c'$ associated with vertex $i$ is marked as \textit{busy} (i.e., being processed by another thread), the current thread skips to the next iteration (line \ref{alg:leidensp--work}). On the other hand, if vertex $i$ has not been visited and community $c'$ is not busy, the thread attempts to mark community $c'$ as \textit{busy} using $atomicCAS()$ (line \ref{alg:leidensp--atomiccas}). If this attempt fails, the thread proceeds to the next iteration. However, if the marking is successful, a BFS is initiated from vertex $i$ to explore the vertices within the same community. This BFS employs lambda functions: $f_{if}$ to selectively execute BFS on vertex $j$ if it belongs to the same community, and $f_{do}$ to update the labels of visited vertices after exploring each vertex during the BFS (line \ref{alg:leidensp--bfs}). Once all vertices have been processed, the threads synchronize, returning the updated labels $C''$, which reflect the new community membership of each vertex without any disconnected communities (line \ref{alg:leidensp--return}). Finally, community $c'$ is marked as \textit{not busy}, allowing it to be processed by other threads. At the end of the algorithm, the updated community membership $C''$ is returned.

\subsubsection{Refinement phase of our Parallel Leiden}

\begin{algorithm}[hbtp]
\caption{Refinement phase of our Parallel Leiden \cite{sahu2024fast}.}
\label{alg:leidenre}
\begin{algorithmic}[1]
\Require{$G'(V', E')$: Input/super-vertex graph}
\Require{$C'_B$: Community bound of each vertex}
\Require{$C'$: Community membership of each vertex}
\Require{$K'$: Total edge weight of each vertex}
\Require{$\Sigma'$: Total edge weight of each community}
\Require{$\Delta R$: Communities marked to be refined}
\Require{$\tau$: Per iteration tolerance}
\Ensure{$G'_{C'}$: Community vertices (CSR)}
\Ensure{$H_t$: Collision-free per-thread hashtable}

\Statex

\Function{leidenRefine}{$G', C'_B, C', K', \Sigma', \Delta R, \tau$} \label{alg:leidenre--move-begin}
  \ForAll{$i \in V'$ \textbf{in parallel}} \label{alg:leidenre--break-marked-begin}
    \If{$\Delta R[C'_B[i]] = 1$} $C'[i] \gets i$ \textbf{;} $\Sigma'[i] \gets K'[i]$
    \EndIf
  \EndFor \label{alg:leidenre--break-marked-end}
  \ForAll{$i \in V'$ \textbf{in parallel}} \label{alg:leidenre--loop-vertices-begin}
    \State $c \gets C'[i]$
    \If{$\Delta R[C'_B[i]] = 0$ \textbf{or} $\Sigma'[c] \neq K'[i]$} \textbf{continue} \label{alg:leidenre--check-isolated}
    \EndIf
    \State $H_t \gets scanBounded(\{\}, G', C'_B, C', i, false)$ \label{alg:leidenre--scan}
    \State $\rhd$ Use $H_t, K', \Sigma'$ to choose best community
    \State $c^* \gets$ Best community linked to $i$ in $G'$ within $C'_B$ \label{alg:leidenre--best-community-begin}
    \State $\delta Q^* \gets$ Delta-modularity of moving $i$ to $c^*$ \label{alg:leidenre--best-community-end}
    \If{$c^* = c$ \textbf{or} $C'[c^*] \neq c^* $} \textbf{continue} \label{alg:leidenre--best-community-same}
    \EndIf
    \If{$atomicCAS(\Sigma'[c], K'[i], 0) = K'[i]$} \label{alg:leidenre--perform-move-begin}
      \State $\Sigma'[c^*] += K'[i]$ \textbf{atomically}
      \State $C'[i] \gets c^*$ \label{alg:leidenre--perform-move-end}
    \EndIf
  \EndFor \label{alg:leidenre--loop-vertices-end}
\EndFunction \label{alg:leidenre--move-end}

\Statex

\Function{scanBounded}{$H_t, G', C'_B, C', i, self$}
  \ForAll{$(j, w) \in G'.edges(i)$}
    \If{\textbf{not} $self$ \textbf{and} $i = j$} \textbf{continue}
    \EndIf
    \If{$C'_B[i] \neq C'_B[j]$} \textbf{continue}
    \EndIf
    \State $H_t[C'[j]] \gets H_t[C'[j]] + w$
  \EndFor
  \Return{$H_t$}
\EndFunction

\Statex

\Function{atomicCAS}{$pointer, old, new$}
  \State $\rhd$ Perform the following atomically
  \If{$pointer = old$} $pointer \gets new$ \textbf{;} \ReturnInline{$old$}
  \Else\ \ReturnInline{$pointer$}
  \EndIf
\EndFunction
\end{algorithmic}
\end{algorithm}

The pseudocode for the refinement phase of our Parallel Leiden algorithm is detailed in Algorithm \ref{alg:leidenlm}. This phase closely resembles the local-moving phase but includes the community membership assigned to each vertex as a \textit{community bound}. During this phase, each vertex must choose a community within its community bound to join, with the goal of maximizing modularity through iterative movements between communities, similar to the local-moving phase. At the start of the refinement phase, the community membership of each vertex is reset, so that each vertex initially represents its own community. The \texttt{leidenRefine()} function is utilized, which takes the current graph $G'$, the community bound for each vertex $C'_B$, the initial community membership $C'$ for each vertex, the total edge weight for each vertex $K'$, the initial total edge weight for each community $\Sigma'$, a flag vector $\Delta R$ that indicates which communities are to be refined, and the current tolerance per iteration $\tau$ as inputs.

The algorithm begins by breaking apart communities that have been marked for refinement in $\Delta R$. It does this by resetting the community memberships of vertices, such that each vertex initially belongs to its own individual community. Following this, we implement the constrained merge procedure, as described in \cite{com-traag19}, during lines \ref{alg:leidenre--loop-vertices-begin}-\ref{alg:leidenre--loop-vertices-end}. This procedure allows vertices within each community boundary to form sub-communities by permitting only isolated vertices (those that belong solely to their own community) to change their community affiliation. This approach effectively divides any internally disconnected communities identified in the local-moving phase and prevents the formation of new disconnected communities. For each isolated vertex $i$ (line \ref{alg:leidenre--check-isolated}), we determine the communities connected to $i$ within the same community boundary, excluding $i$ itself (line \ref{alg:leidenre--scan}). The refinement phase is skipped for a vertex if its community has not been marked for refinement ($\Delta R[C'_B[i]] = 0$) or if another vertex has joined its community, which is indicated by the total community weight no longer aligning with the total edge weight of the vertex. Next, we identify the optimal community $c^*$ for relocating $i$ (line \ref{alg:leidenre--best-community-begin}) and evaluate the delta-modularity of moving $i$ to $c^*$ (line \ref{alg:leidenre--best-community-end}). If a better community is found, we attempt to update the community membership of $i$, provided that it is still isolated (lines \ref{alg:leidenre--perform-move-begin}-\ref{alg:leidenre--perform-move-end}). We do not migrate vertex $i$ to community $c^*$ if the vertex representing $c^*$ has moved to a different community.

\subsubsection{Aggregation phase of our Parallel Leiden}

The pseudocode for the aggregation phase is detailed in Algorithm \ref{alg:leidenag}, where communities are combined into super-vertices. In particular, the \texttt{leidenAggre} \texttt{gate()} function of this algorithm accepts the current graph $G'$ and the community membership $C'$ of vertices in the current pass as inputs, and returns the super-vertex graph $G''$ as output.

In the algorithm, the process begins by obtaining the offsets array for the community vertices in the CSR format, referred to as $G'_{C'}.offsets$, as detailed in lines \ref{alg:leidenag--coff-begin} to \ref{alg:leidenag--coff-end}. This involves counting the number of vertices in each community using the function \texttt{countCommunityVertices()}, followed by performing an exclusive scan on the resulting array. Next, in lines \ref{alg:leidenag--comv-begin} to \ref{alg:leidenag--comv-end}, we concurrently traverse all vertices and atomically assign the vertices associated with each community into the community graph CSR $G'_{C'}$. Subsequently, the offsets array for the super-vertex graph CSR is computed by estimating the degree of each super-vertex in lines \ref{alg:leidenag--yoff-begin} to \ref{alg:leidenag--yoff-end}. This includes calculating the total degree of each community using \texttt{communityTotalDegree()} and executing another exclusive scan. As a result, the super-vertex graph CSR is structured with intervals for the edges and weights array of each super-vertex. Then, in lines \ref{alg:leidenag--y-begin} to \ref{alg:leidenag--y-end}, we iterate over all communities $c \in [0, |\Gamma|)$ in parallel, utilizing dynamic loop scheduling with a chunk size of $2048$ for Static Leiden, and a chunk size of $32$ for ND, DS, and DF Leiden. During this stage, all communities $d$ (along with their corresponding edge weights $w$) connected to each vertex $i$ in community $c$ are included (via \texttt{scanCommunities()}, as described in Algorithm \ref{alg:leidenlm}) in the per-thread hashtable $H_t$. Once $H_t$ contains all connected communities and their weights, they are atomically added as edges to super-vertex $c$ in the super-vertex graph $G''$. Finally, in line \ref{alg:leidenag--return}, we return the super-vertex graph $G''$.

\begin{algorithm}[hbtp]
\caption{Aggregation phase of our Parallel Leiden \cite{sahu2024fast}.}
\label{alg:leidenag}
\begin{algorithmic}[1]
\Require{$G'(V', E')$: Input/super-vertex graph}
\Require{$C'$: Community membership of each vertex}
\Ensure{$G'_{C'}$: Community vertices (CSR)}
\Ensure{$G''$: Super-vertex graph (weighted CSR)}
\Ensure{$*.offsets$: Offsets array of a CSR graph}
\Ensure{$H_t$: Collision-free per-thread hashtable}

\Statex

\Function{leidenAggregate}{$G', C'$}
  \State $\rhd$ Obtain vertices belonging to each community
  \State $G'_{C'}.offsets \gets countCommunityVertices(G', C')$ \label{alg:leidenag--coff-begin}
  \State $G'_{C'}.offsets \gets exclusiveScan(G'_{C'}.offsets)$ \label{alg:leidenag--coff-end}
  \ForAll{$i \in V'$ \textbf{in parallel}} \label{alg:leidenag--comv-begin}
    \State Add edge $(C'[i], i)$ to CSR $G'_{C'}$ atomically
  \EndFor \label{alg:leidenag--comv-end}
  \State $\rhd$ Obtain super-vertex graph
  \State $G''.offsets \gets communityTotalDegree(G', C')$ \label{alg:leidenag--yoff-begin}
  \State $G''.offsets \gets exclusiveScan(G''.offsets)$ \label{alg:leidenag--yoff-end}
  \State $|\Gamma| \gets$ Number of communities in $C'$
  \ForAll{$c \in [0, |\Gamma|)$ \textbf{in parallel}} \label{alg:leidenag--y-begin}
    \If{degree of $c$ in $G'_{C'} = 0$} \textbf{continue}
    \EndIf
    \State $H_t \gets \{\}$
    \ForAll{$i \in G'_{C'}.edges(c)$}
      \State $H_t \gets scanCommunities(H, G', C', i, true)$
    \EndFor
    \ForAll{$(d, w) \in H_t$}
      \State Add edge $(c, d, w)$ to CSR $G''$ atomically
    \EndFor
  \EndFor \label{alg:leidenag--y-end}
  \Return $G''$ \label{alg:leidenag--return}
\EndFunction
\end{algorithmic}
\end{algorithm}

\subsection{Marking communities for Selective splitting and refinement}

Algorithm \ref{alg:leidenmk} is designed to identify communities that require splitting or refining, based on recent edge updates. It takes as input the current graph $G^t$, batches of edge deletions $\Delta^{t-}$ and insertions $\Delta^{t+}$, the previous community assignments $C^{t-1}$ of each vertex, the current total edge weights $\Sigma^t$ of communities, and the change in total edge weights $\Delta \Sigma^t$ of communities.

The algorithm starts by initializing two flag vectors, $\Delta R$ and $\Delta S$, to hold the communities that are marked for refinement and splitting, respectively. If the algorithm is \textit{static}, all vertices in the graph are marked for refinement. However, if the algorithm is dynamic, it first processes the edge deletions in parallel, checking whether the vertices $i$ and $j$ belong to the same community. If they do, it marks the community in $\Delta S$. Subsequently, the algorithm checks both edge deletions and insertions to determine if any communities should be refined. It does this by examining the edge weight changes; if two vertices belong to the same community and the relative change in total edge weight for that community exceeds a specified refinement tolerance $\tau_{re}$, it marks the community for refinement in $\Delta R$. Finally, $\Delta S$ and $\Delta R$ are returned.

\subsection{Renumbering communities by ID of a vertex within}

We will now present Algorithm \ref{alg:leidensr}, which details a method for renumbering communities according to their internal vertices. The objective is to assign each community an ID corresponding to one of its member vertices. The algorithm takes several inputs: the current graph $G'$, the vertex community assignments $C'$, the total edge weights of the communities $\Sigma'$, as well as the communities designated for splitting $\Delta S$ and those marked for refinement $\Delta R$.

The algorithm starts by initializing several key data structures (line \ref{alg:leidensr--init}). It then iterates over each vertex $i$ to retrieve the current community ID $c'$ from $C'[i]$. If no representative vertex has been assigned for community $c'$ (i.e., $C'_v[c']$ is empty), vertex $i$ is designated as the representative for that community. Next, the second parallel loop processes all communities $c' \in \Gamma'$, which is the set of communities in the original graph. For each community $c'$, the representative vertex $c''$ is obtained from the previous step, and the total edge weight $\Sigma'[c']$ along with the flags $\Delta S[c']$ and $\Delta R[c']$ are reorganized into their updated versions: $\Sigma''[c'']$, $\Delta S'[c'']$, and $\Delta R'[c'']$. Finally, the third parallel loop updates the community membership for each vertex based on the representative vertex assigned to its community. For each vertex $i$, the algorithm determines the current community $C'[i]$, retrieves the representative vertex for that community, and reassigns vertex $i$ to the new community ID that corresponds to its representative. After updating the community memberships, the algorithm performs an in-place update of the original structures: the community memberships $C'$, total edge weights $\Sigma'$, and flags $\Delta S$ and $\Delta R$ are replaced with the updated versions $C''$, $\Sigma''$, $\Delta S'$, and $\Delta R'$.

\begin{algorithm}[hbtp]
\caption{Renumber communities by ID of a vertex within.}
\label{alg:leidensr}
\begin{algorithmic}[1]
\Require{$G'(V', E')$: Input/super-vertex graph}
\Require{$C', C''$: Current, updated community membership of vertices}
\Require{$\Sigma', \Sigma''$: Current, updated total edge weight of each community}
\Require{$\Delta S, \Delta S'$: Current, updated communities marked for splitting}
\Require{$\Delta R, \Delta R'$: Current, updated communities marked for refining}
\Ensure{$\Gamma'$: Set of communities in $C'$}

\Statex

\Function{leidenSubsetRenumber}{$G', C', \Sigma', \Delta S, \Delta R$}
  \State $C'' \gets \Sigma'' \gets \Delta S' \gets \Delta R' \gets C'_v \gets \{\}$ \label{alg:leidensr--init}
  \State $\rhd$ Obtain any vertex from each community
  \ForAll{$i \in V'$ \textbf{in parallel}}
    \State $c' \gets C'[i]$
    \If{$C'_v[c] = \text{EMPTY}$} $C'_v[c] \gets i$
    \EndIf
  \EndFor
  \State $\rhd$ Update community weights and changed status
  \ForAll{$c' \in \Gamma'$ \textbf{in parallel}}
    \State $c'' \gets C'_v[c']$
    \If{$c'' \neq \text{EMPTY}$}
      \State $\Sigma''[c''] \gets \Sigma'[c']$
      \State $\Delta S'[c''] \gets \Delta S[c']$
      \State $\Delta R'[c''] \gets \Delta R[c']$
    \EndIf
  \EndFor
  \State $\rhd$ Update community memberships
  \ForAll{$i \in V'$ \textbf{in parallel}}
    \State $C''[i] \gets C'_v[C'[i]]$
  \EndFor
  \State $\rhd$ Update in-place
  \State $C' \gets C''$ \textbf{;} $\Sigma' \gets \Sigma''$ \textbf{;} $\Delta S \gets \Delta S'$ \textbf{;} $\Delta R \gets \Delta R'$
\EndFunction
\end{algorithmic}
\end{algorithm}

\subsection{Updating vertex/community weights}
\label{sec:our-update}

We will now detail the parallel algorithm developed to compute the updated weighted degree of each vertex $K^t$ and the total edge weight of each community $\Sigma^t$, \texttt{updateWeights()}. It relies on the previous community memberships of the vertices $C^{t-1}$, their weighted degrees $K^{t-1}$, the total edge weights of the communities, and a batch update that includes edge deletions $\Delta^{t-}$ and insertions $\Delta^{t+}$. The pseudocode for this can be found in Algorithm \ref{alg:update}.

The algorithm begins with the initialization of $K$ and $\Sigma$, which represent the weighted degree of each vertex and the total edge weight of each community, respectively (line \ref{alg:update--init}). Following this, we utilize multiple threads to process sets of edge deletions $\Delta^{t-}$ (lines \ref{alg:update--loopdel-begin}-\ref{alg:update--loopdel-end}) and edge insertions $\Delta^{t+}$ (lines \ref{alg:update--loopins-begin}-\ref{alg:update--loopins-end}). For each edge deletion $(i, j, w)$ in $\Delta^{t-}$, we identify the community $c$ of vertex $i$ based on the previous community assignment $C^{t-1}$ (line \ref{alg:update--delc}). If vertex $i$ is included in the current thread's work-list, we decrement its weighted degree by $w$ (line \ref{alg:update--delk}). Additionally, if community $c$ is part of the work-list, its total edge weight is also reduced by $w$ (line \ref{alg:update--delsigma}). Likewise, for each edge insertion $(i, j, w)$ in $\Delta^{t+}$, we adjust the weighted degree of vertex $i$ and the total edge weight of its community. Finally, we return the updated values of $K$ and $\Sigma$ for each vertex and community for further processing (line \ref{alg:update--return}).

\begin{algorithm}[hbtp]
\caption{Updating vertex/community weights in parallel\ignore{\cite{sahu2024dflouvain}}.}
\label{alg:update}
\begin{algorithmic}[1]
\Require{$G^t(V^t, E^t)$: Current input graph}
\Require{$\Delta^{t-}, \Delta^{t+}$: Edge deletions and insertions (batch update)}
\Require{$C^{t-1}$: Previous community of each vertex}
\Require{$K^{t-1}$: Previous weighted-degree of each vertex}
\Require{$\Sigma^{t-1}$: Previous total edge weight of each community}
\Require{$\Delta \Sigma^{t-1}$: Previous change in total edge weight of communities}
\Ensure{$K$: Updated weighted-degree of each vertex}
\Ensure{$\Sigma$: Updated total edge weight of each community}
\Ensure{$\Delta \Sigma$: Updated change in total edge weight of communities}
\Ensure{$work_{th}$: Work-list of current thread}

\Statex

\Function{updateWeights}{$G^t, \Delta^{t-}, \Delta^{t+}, C^{t-1}, K^{t-1}, \Sigma^{t-1}$}
  \State $K \gets K^{t-1}$ \textbf{;} $\Sigma \gets \Sigma^{t-1}$ \label{alg:update--init}
  \ForAll{\textbf{threads in parallel}} \label{alg:update--loopdel-begin}
    \ForAll{$(i, j, w) \in \Delta^{t-}$}
      \State $c \gets C^{t-1}[i]$ \label{alg:update--delc}
      \If{$i \in work_{th}$} $K[i] \gets K[i] - w$ \label{alg:update--delk}
      \EndIf
      \If{$c \in work_{th}$} $\Sigma[c] \gets \Sigma[c] - w$ \label{alg:update--delsigma}
      \EndIf
    \EndFor \label{alg:update--loopdel-end}
    \ForAll{$(i, j, w) \in \Delta^{t+}$} \label{alg:update--loopins-begin}
      \State $c \gets C^{t-1}[i]$
      \If{$i \in work_{th}$} $K[i] \gets K[i] + w$
      \EndIf
      \If{$c \in work_{th}$} $\Sigma[c] \gets \Sigma[c] + w$
      \EndIf
    \EndFor
  \EndFor \label{alg:update--loopins-end}
  \Return $\{K, \Sigma\}$ \label{alg:update--return}
\EndFunction

\Statex

\Function{updateChanges}{$G^t, \Delta^{t-}, \Delta^{t+}, C^{t-1}, \Delta \Sigma^{t-1}$}
  \State $\Delta \Sigma \gets \Delta \Sigma^{t-1}$ \label{alg:updatech--init}
  \ForAll{\textbf{threads in parallel}} \label{alg:updatech--loop-begin}
    \ForAll{$(i, j, w) \in \Delta^{t-} \cup \Delta^{t+}$}
      \State $c \gets C^{t-1}[i]$ \label{alg:updatech--delc}
      \If{$c \in work_{th}$} $\Delta \Sigma[c] \gets \Delta \Sigma[c] + w$ \label{alg:updatech--delsigma}
      \EndIf
    \EndFor
  \EndFor \label{alg:updatech--loop-end}
  \Return $\Delta \Sigma$ \label{alg:updatech--return}
\EndFunction
\end{algorithmic}
\end{algorithm}

In Algorithm \ref{alg:update}, the \texttt{updateChanges()} function serves a purpose analogous to that of \texttt{updateWeights()}. It calculates the revised cumulative change in the total edge weight for each community, denoted as $\Delta \Sigma$, based on the previous cumulative change in total edge weight for each community, the previous community memberships of the vertices $C^{t-1}$, and a batch update that incorporates edge deletions $\Delta^{t-}$ and insertions $\Delta^{t+}$.

The algorithm starts by initializing $\Delta \Sigma$. Next, we employ multiple threads to handle edge deletions $\Delta^{t-}$ and edge insertions $\Delta^{t+}$. For each edge deletion or insertion $(i, j, w)$ in the set $\Delta^{t-} \cup \Delta^{t+}$, we determine the community $c$ of vertex $i$ using the previous community assignment $C^{t-1}$. If vertex $i$ is part of the current thread's work list, we increase the total edge weight change of community $c$ by $w$. We assume that the batch updates are undirected, which means the change in community weight for the other end of the edge will be processed automatically. Finally, we return the updated $\Delta \Sigma$ for each community for further processing.

\subsection{Renumbering communities to enable tracking over time}
\label{sec:our-tracking}

The algorithm outlined in Algorithm \ref{alg:leidentr} aims to renumber communities within a dynamic graph to facilitate tracking their evolution over time. It takes as input the current graph snapshot $G^t$, the current and previous community assignments $C^t$ and $C^{t-1}$, and the current weighted-degree $K^t$ of each vertex.

The algorithm starts by initializing several structures: $H$ for storing the most overlapping community for each old community, $H'$ for mapping the top old community ID with overlap weight, and $C'$ for updating the community IDs of each vertex. Next, it iterates over all vertices $i$ in parallel to identify overlapping communities. For each vertex, it retrieves the previous community $c$ from $C^{t-1}$, the current community $c'$ from $C^t$, and the weighted degree $w$ from $K^t$. If the current community $c'$ matches the most overlapping community $H_k[c]$ for the old community $c$, it updates the overlap weight $H_v[c]$ by adding $w$. If $H_v[c]$ is greater than $w$, it subtracts $w$ from $H_v[c]$. If neither condition is met, it assigns $H[c]$ to the pair $\{c', w\}$. The algorithm then processes each key in $H$ in parallel to determine the best old community. For each community $c$ in $H$, it retrieves the old community $c'$ and the corresponding overlap weight $w$. If the overlap weight for the old community $H'_v[c']$ is less than or equal to $w$, it updates $H'[c']$ to $\{c, w\}$. The algorithm assigns new community IDs to unassigned communities by generating a random community ID $c$ within the range of $1$ to the number of vertices $|V^t|$ on each thread. For each community $c'$ in the set of current communities $\Gamma^t$, it continues if the old community count $H'_k[c']$ is not zero. It enters a loop where it increments the community ID $c$ until it finds an ID that is unassigned (where $H_k[c]$ is empty). It uses an atomic compare-and-swap operation to ensure safe assignment of community IDs in a concurrent environment. Finally, the algorithm assigns the new community IDs to each vertex by iterating over all vertices $i$ in parallel and updating $C'[i]$ with the newly assigned ID from $H'_k[C^t[i]]$. The function concludes by returning the updated community ID assignment $C'$.

\begin{algorithm}[hbtp]
\caption{Renumber communities to enable tracking\ignore{over time}.}
\label{alg:leidentr}
\begin{algorithmic}[1]
\Require{$G^t(V^t, E^t)$: Current input graph}
\Require{$C^t, C^{t-1}$: Current, previous community of each vertex}
\Require{$K^t$: Current weighted-degree of each vertex}
\Ensure{$H$: Top community with overlap weight for each old one}
\Ensure{$H'$: Top old ID with overlap weight for each community}
\Ensure{$C'$: Updated community ID of each vertex}
\Ensure{$\Gamma^t$: Set of communities in $C^t$}

\Statex

\Function{leidenTrack}{$G^t, C^t, C^{t-1}, K^t$}
  \State $H \gets H' \gets C' \gets \{\}$
  \State $\rhd$ Most overlapping community for each old community
  \ForAll{$i \in V^t$ \textbf{in parallel}}
    \State $c \gets C^{t-1}[i]$ \textbf{;} $c' \gets C^t[i]$ \textbf{;} $w \gets K^t[i]$
    \If{$H_k[c] = c'$} $H_v[c]\ \text{+=}\ w$
    \ElsIf{$H_v[c] > w$} $H_v[c]\ \text{-=}\ w$
    \Else\ $H[c] \gets \{c', w\}$ 
    \EndIf
  \EndFor
  \State $\rhd$ Find best old community for each community
  \ForAll{$c \in H.keys()$ \textbf{in parallel}}
    \State $c' \gets H_k[c]$ \textbf{;} $w \gets H_v[c]$
    \If{$H'_v[c'] \leq w$} $H'[c'] \gets \{c, w\}$
    \EndIf
  \EndFor
  \State $\rhd$ Assign a free community ID to unassigned communities
  \State $c \gets \textbf{random in}\ [1, |V^t|)$ \textbf{on each thread}
  \ForAll{$c' \in \Gamma^t$ \textbf{in parallel}}
    \If{$H'_k[c'] \neq 0$} \textbf{continue}
    \EndIf
    \Loop
      \While{$H_k[c] \neq \emptyset$} $c \gets (c + 1)\ \text{mod}\ |V^t|$
      \EndWhile
      \If{$atomicCAS(H_k[c], \emptyset, c')$ \textbf{passes}} \textbf{break}
      \EndIf
    \EndLoop
    \State $H'[c'] \gets c$
  \EndFor
  \State $\rhd$ Assign new community IDs
  \ForAll{$i \in V^t$ \textbf{in parallel}}
    \State $C'[i] \gets H'_k[C^t[i]]$
  \EndFor
  \Return{$C'$}
\EndFunction
\end{algorithmic}
\end{algorithm}

\subsection{Performance Comparison on Real-world dynamic graphs}
\label{sec:performance-comparison-temporal}

\begin{figure*}[!hbt]
  \centering
  \subfigure[Overall Runtime]{
    \label{fig:temporal-summary--runtime-overall}
    \includegraphics[width=0.48\linewidth]{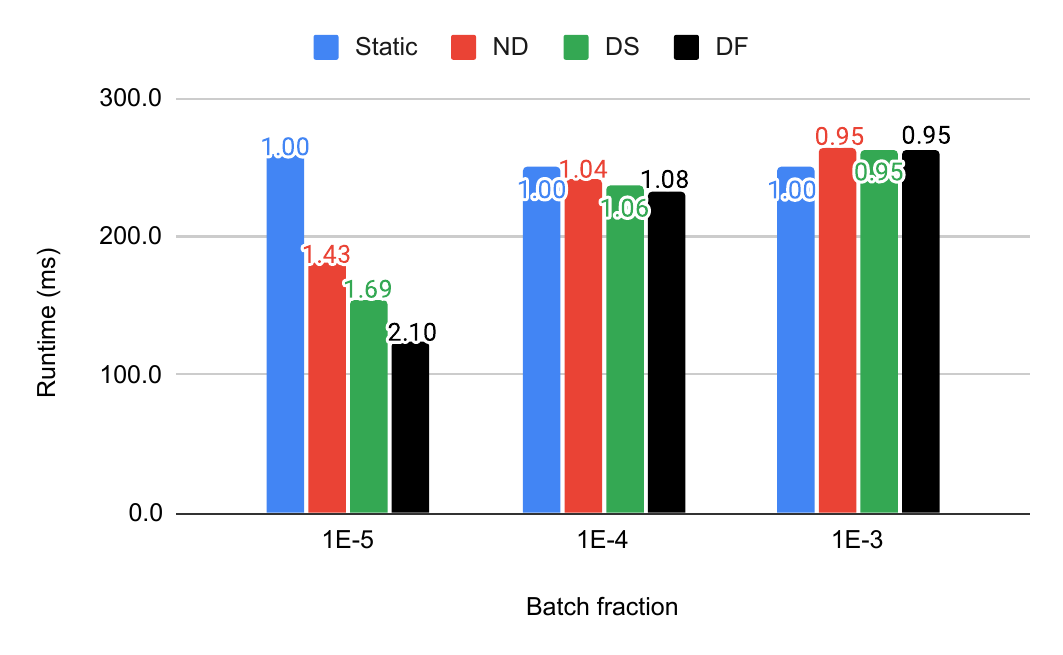}
  }
  \subfigure[Overall Modularity of communities obtained]{
    \label{fig:temporal-summary--modularity-overall}
    \includegraphics[width=0.48\linewidth]{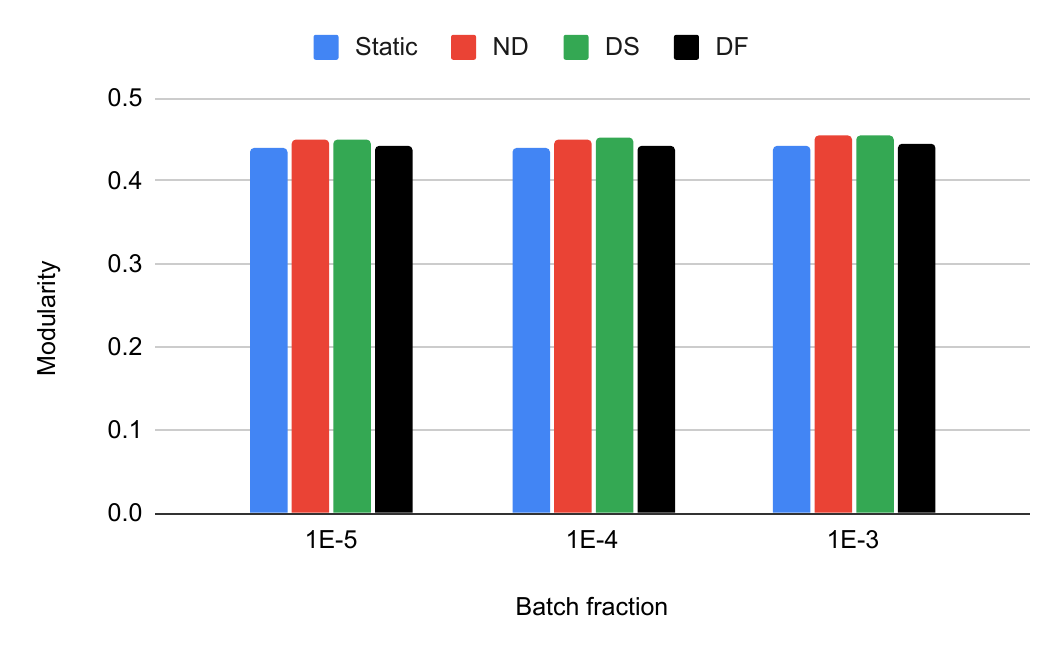}
  } \\[2ex]
  \includegraphics[width=0.48\linewidth]{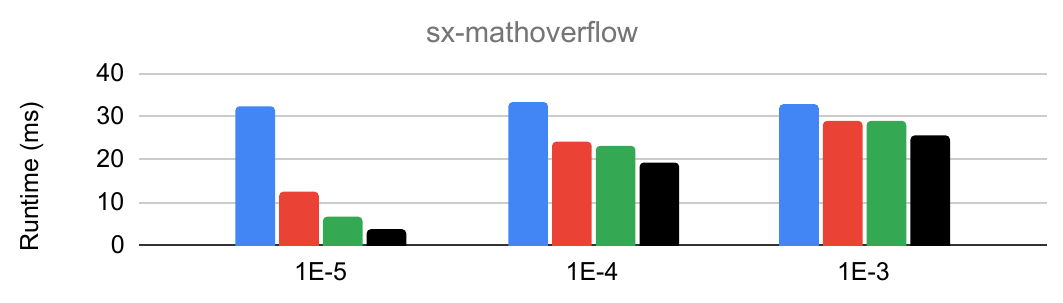}
  \includegraphics[width=0.48\linewidth]{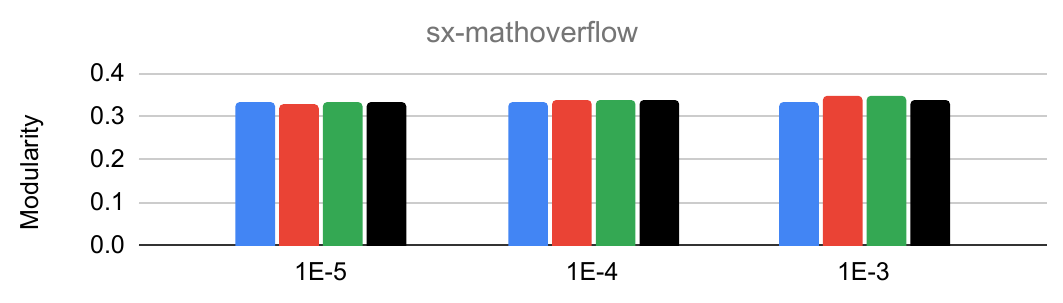}
  \includegraphics[width=0.48\linewidth]{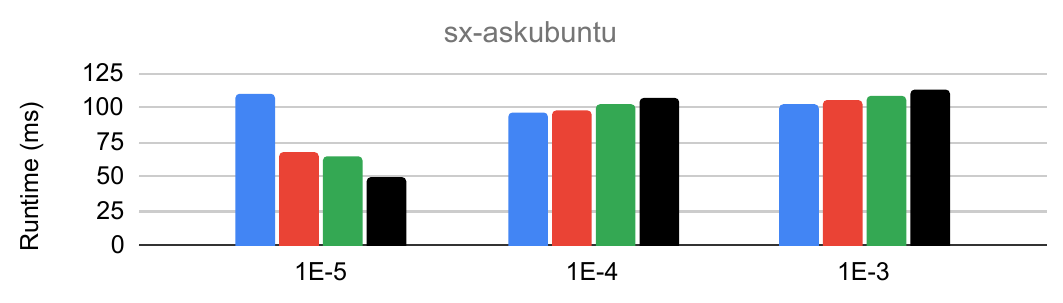}
  \includegraphics[width=0.48\linewidth]{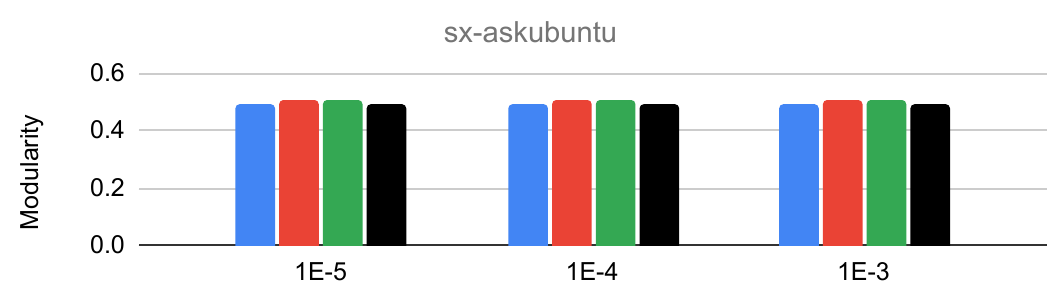}
  \includegraphics[width=0.48\linewidth]{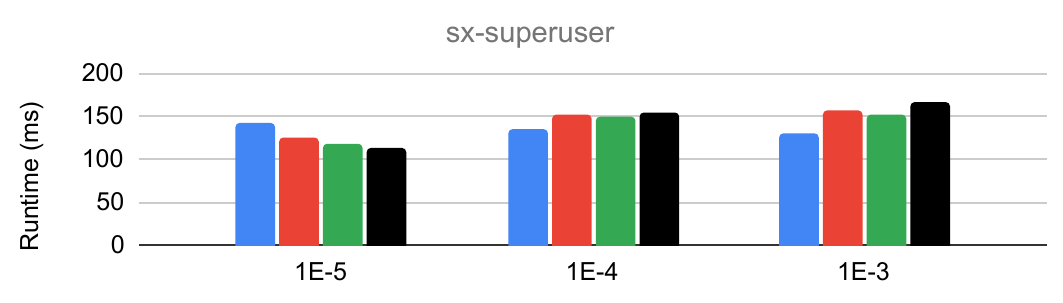}
  \includegraphics[width=0.48\linewidth]{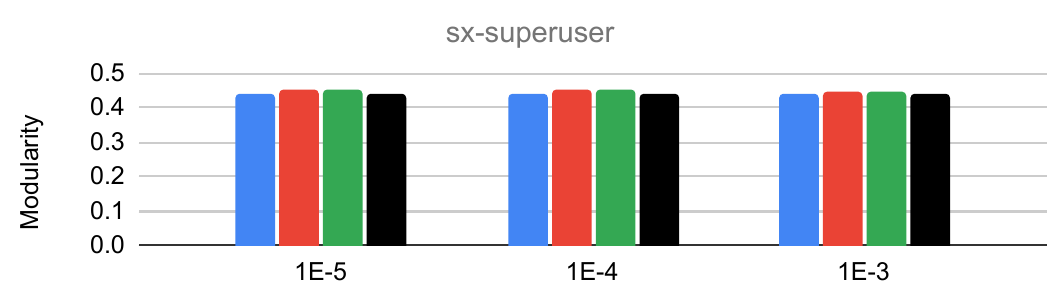}
  \includegraphics[width=0.48\linewidth]{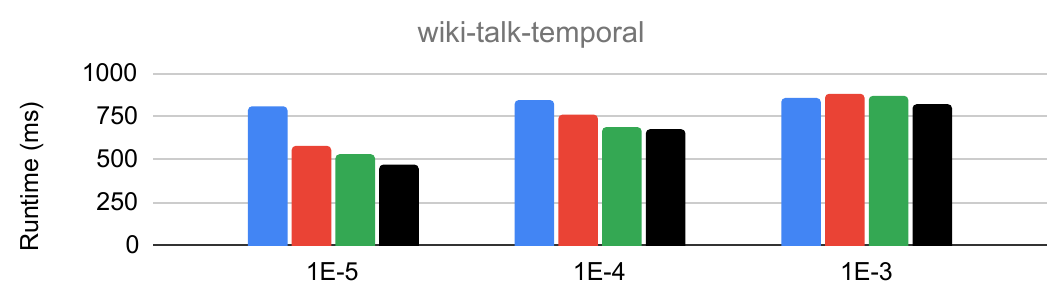}
  \includegraphics[width=0.48\linewidth]{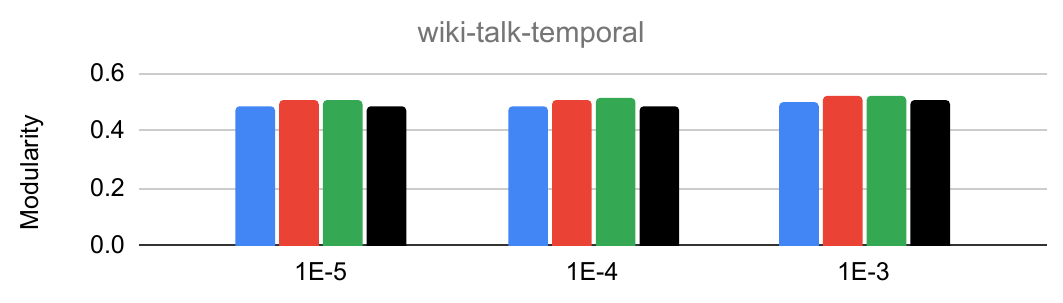}
  \subfigure[Runtime on each dynamic graph]{
    \label{fig:temporal-summary--runtime-graph}
    \includegraphics[width=0.48\linewidth]{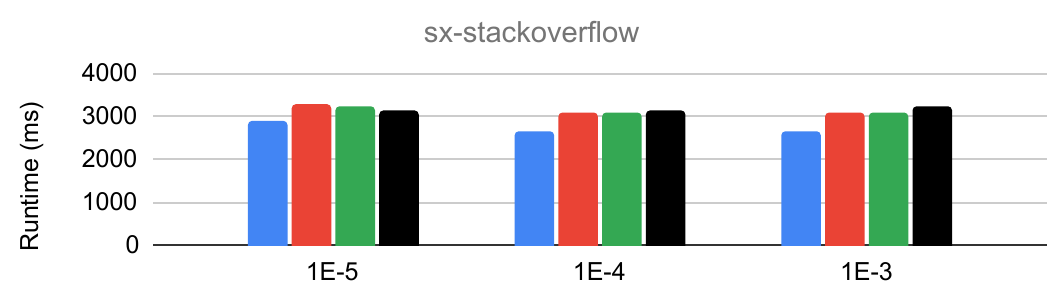}
  }
  \subfigure[Modularity in communities obtained on each dynamic graph]{
    \label{fig:temporal-summary--modularity-graph}
    \includegraphics[width=0.48\linewidth]{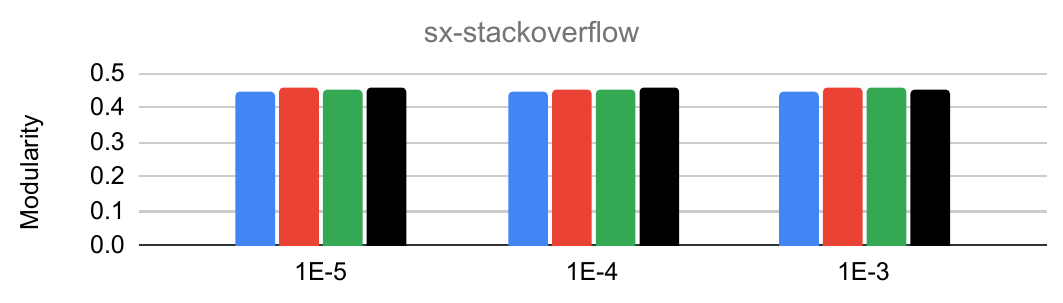}
  } \\[-2ex]
  \caption{Mean Runtime and Modularity of communities obtained with parallel \textit{Static}, and our improved \textit{Naive-dynamic (ND)}, \textit{Delta-screening (DS)}, and \textit{Dynamic Frontier (DF) Leiden} on real-world dynamic graphs, with batch updates of size $10^{-5}|E_T|$ to $10^{-3}|E_T|$. Here, (a) and (b) show the overall runtime and modularity across all temporal graphs, while (c) and (d) present the runtime and modularity for each individual graph. In (a), the speedup of each approach relative to Static Leiden is labeled.}
  \label{fig:temporal-summary}
\end{figure*}

We also assess the performance of parallel implementations of Static, along with our improved ND, DS, and DF Leiden, on the real-world dynamic graphs listed in Table \ref{tab:dataset}. These evaluations are conducted on batch updates of size $10^{-5}|E_T|$ to $10^{-3}|E_T|$. For each batch size, as outlined in Section \ref{sec:batch-generation}, we load $90\%$ of the graph, add reverse edges to ensure undirected edges, and then load $B$ edges (where $B$ is the batch size) across $100$ consecutive batch updates. Figure \ref{fig:temporal-summary--runtime-overall} illustrates the overall runtime for each method across all graphs and batch sizes, while Figure \ref{fig:temporal-summary--modularity-overall} shows the overall modularity of the resulting communities. Additionally, Figures \ref{fig:temporal-summary--runtime-graph} and \ref{fig:temporal-summary--modularity-graph} display the average runtime and modularity achieved by each method for the individual dynamic graphs in the dataset.

Figure \ref{fig:temporal-summary--runtime-overall} shows that, on average, ND, DS, and DF Leiden are $1.14\times$, $1.23\times$, and $1.38\times$ faster than Static Leiden for batch updates of size $10^{-5}|E_T|$ to $10^{-3}|E_T|$. We now discuss why these methods provide only modest speed improvements over Static Leiden. Our experiments reveal that only around $30\%$ of the total runtime for ND, DS, and DF Leiden is spent in the first pass of the algorithm. Additionally, for large batch updates, a significant portion of the runtime is dedicated to the splitting, refinement, and aggregation phases. For smaller batch updates, a substantial amount of time is spent renumbering communities. As a result, the overall speedup of ND, DS, and DF Leiden compared to Static Leiden remains limited.

\end{document}